\documentclass[12pt]{article}

\usepackage{url}

\usepackage{graphicx}
\usepackage{color}    
\usepackage{cite}
\usepackage{amsmath}
\usepackage{amssymb}
\usepackage{lineno,hyperref}
\modulolinenumbers[5]

\def\beq{\begin{equation}}
\def\eeq{\end{equation}}
\def\nn{\nonumber}
\def\bea{\begin{eqnarray}}
\def\eea{\end{eqnarray}}
\def\ba{\begin{array}}                  
\def\ea{\end{array}}

\evensidemargin \oddsidemargin
\begin{document}

\thispagestyle{empty}

\def\thefootnote{\fnsymbol{footnote}}


\vspace{1cm}

\begin{center}
{\large\sc {\bf $Q_{6}$ As The Flavour Symmetry in a  Non-minimal SUSY $SU(5)$ Model}}

\vspace{1.4cm}

{\sc 
 J. C. G\'omez-Izquierdo$^{1}$\footnote{email: jcarlos@fisica.unam.mx}, 
 F. Gonz\'alez-Canales$^{2,3}$\footnote{email: felix.gonzalez@ific.uv.es} and  
 M~Mondragon$^{1}$\footnote{email: myriam@fisica.unam.mx}
}

\vspace*{1cm}

{\sl 
 $^{1}$~Instituto~de F{\'{\i}}sica, Universidad~Nacional Aut\'onoma de M\'exico\\ 
  M\'exico 01000, D.F., M\'exico \\
 $^{2}$~Facultad de Ciencias de la Electr\'onica, Benem\'erita Universidad Aut\'onoma \\
  de Puebla,  Apdo. Postal 157, 72570, Puebla, Pue., M\'exico \\ 
 $^{3}$~AHEP Group, Instituto de F\'{\i}sica Corpuscular-CSIC/Universitat de Valencia, \\
  Parc Cientific de Paterna, C/ Catedr\'atico Jos\'e Beltr\'an, 2, 
  E-46980 Paterna (Valencia) Spain}

\end{center}

\vspace*{0.2cm}

\begin{abstract}
  We present a non-minimal renormalizable SUSY $SU(5)$~model, with
  extended Higgs sector and right-handed neutrinos, where the flavour
  sector exhibits a $Q_{6}$ flavour symmetry. We analysed the simplest
  version of this model, in which R-parity is conserved and the
  right-handed neutrino masses in the flavour doublet are considered
  with and without degeneration. We find the generic form of the mass
  matrices both in the quark and lepton sectors. We reproduce,
  according to current data, the mixing in the CKM matrix.  In the
  leptonic sector, in the general case where the right-handed
  neutrino masses are not degenerate, we find that the values for the
  solar, atmospheric, and reactor mixing angles are in very good 
  agreement with the experimental data, both for a normal and an 
  inverted hierarchy.  In the particular case where the right handed 
  neutrinos masses are degenerate, the model predicts a strong inverted 
  hierarchy spectrum and a sum rule among the neutrino masses. In this case the
  atmospheric and solar angles are in very good agreement with
  experimental data, and the reactor one is different from zero,
  albeit too small ($\theta^{\ell^{th}}_{13} \sim 3.38^{\circ} $). This value
  constitues a lower bound for $\theta_{13}$ in the general case.  We
  also find the range of values for the neutrino masses in each case.
  
\end{abstract}

\def\thefootnote{\arabic{footnote}}
\setcounter{page}{0}
\setcounter{footnote}{0}
\newpage


\section{Introduction}
 Understanding the flavour sector of the Standard Model (SM) has been a
 puzzle for a long time, due to the large differences in the Yukawa
 couplings of the different fermions. The hierarchy between the fundamental
 particles, the amount of CP violation and the structure of the CKM matrix 
 remains as open~\cite{Masiero:2005ua}. In spite of these subtle facts, the
 success of the SM is remarkable.

 One of the strategies to deal with the flavour problem in the quark
 and lepton sectors has been to study it in the framework of textures
 (zeros) in the mass matrices. The textures have been explored for a
 long time,  as an attempt to eliminate the irrelevant free
 parameters in the Yukawa sector. As it is well known, the Fritzsch
 textures can accommodate the quarks mixing angles, in terms of the quarks 
 masses~\cite{Fritzsch:1977za, Fritzsch:1977vd}. This approach
 seems to work out correctly because the Cabbibo angle is obtained with
 great accuracy~\cite{Fritzsch:1977za}. However, this framework
 presents some problems with the top mass and the $V_{cb}$ element of
 the CKM matrix~\cite{Kobayashi:1973fv, Cabibbo:1963yz}, as can be
 seen in~\cite{Branco:2010tx, Fritzsch:2011cu}. Recently, deviations to
 the Fritzsch textures have appeared in order to overcome these
 problems. Moreover, the charged lepton and neutrino sector have been
 included in this kind of ansatz and consistent results on the PMNS
 matrix~\cite{Maki:1962mu,Pontecorvo:1967fh} have been observed in this
 generic approximation~\cite{Fritzsch:2011cu}. As alternative textures
 to the Fritzsch ones, the Nearest Neighbour Interaction
 (NNI) textures can also reproduce very well the flavour mixing in the 
 quark and lepton sectors~\cite{Harayama:1996am, Harayama:1996jr}; it is 
 well known that Fritzsch textures could be obtained from the NNI ones as
 a limiting case~\cite{Harayama:1996am}.

 Non-Abelian flavour symmetries have been playing an important role in
 model building, these symmetries are considered as an elegant way to
 obtain the NNI textures in the fermion mass
 matrices~\cite{Harayama:1996jr, Harayama:1996am}.  In particular, the
 flavour symmetry group $Q_{6}$~\cite{Ishimori:2010au} has been
 proposed as responsible of these kind of textures in the quarks as
 well as in the leptonic sector~\cite{Babu:2004tn, Kajiyama:2005rk,
   Kajiyama:2007pr, Babu:2009nn, Babu:2011mv}. The rich phenomenology
 that $Q_{6}$ provides in the SUSY scenario is remarkable, one of
 these features is to prohibit the dangerous terms that mediate fast
 proton decay, rather than invoking the $R$-parity
 symmetry~\cite{Kajiyama:2005rk,Kajiyama:2007pr}. The immediate
 question that arises is, how does the $Q_{6}$ symmetry work within a GUT
 framework? In particular, the main question that we will address here
 is if the SUSY-$SU(5)$ models are compatible with the $Q_{6}$ group
 in order to accommodate masses and mixings for fermions.

 From a theoretical point of view, Grand Unified ideas are well
 motivated for fundamental reasons~\cite{Georgi:1974yf,Pati:1974yy,
 Mohapatra:1974gc,Georgi:1974sy}. In particular, the $SU(5)$
 model~\cite{Georgi:1974sy} is considered to be one of the best
 scenarios to unify the electroweak and strong interactions. However,
 the model itself faces serious phenomenological and theoretical
 problems, one of them being that active neutrinos are
 massless~\cite{Langacker:1980js,Buras:1977yy}; another one that the
 unification of gauge couplings is not quite good with the current
 precision data. The simplicity of the $SU(5)$ can be retained even if
 it is promoted to be a supersymmetric model. The SUSY
 $SU(5)$~\cite{Dimopoulos:1981zb, Dimopoulos:1981yj, Sakai:1981gr}
 version, with $R$-parity conserved, gives a much better unification
 of the gauge couplings, but new couplings (five dimensional
 operators~\cite{Sakai:1981pk, Weinberg:1981wj}) can yield the proton
 decay~\cite{Ellis:1981tv, Dimopoulos:1981dw, Nath:1985ub}, and
 therefore, they can exclude the SUSY version~\cite{Hisano:1992jj,
   Goto:1998qg, Murayama:2001ur}.  On the other hand, generic studies
 on the minimal SUSY $SU(5)$ model have made clear that it may not be
 ruled out by the experimental data on the proton decay
 rates~\cite{Bajc:2002bv,Martens:2010nm,Hisano:2013exa}. Taking into
 account the neutrino mass problem, the SUSY $SU(5)$ version (SUSY
 models in general) provides elegant mechanisms to generate massive
 neutrinos via $R$-parity violation~\cite{Romao:1991ex,
   Hirsch:2000ef}. But still, the simplest way to give mass to the
 left-handed neutrinos in the minimal $SU(5)$ or SUSY scenario is to
 consider three right-handed neutrinos (RHN's) which are singlets
 under the gauge group and invoke the type I see-saw
 mechanism~\cite{GellMann:1980vs,fukugita2003physics,Mohapatra:1980yp,Mohapatra:1979ia,Minkowski:1977sc}.

 There are interesting SUSY $SU(5)$ 
 models~\cite{Ishimori:2008fi,Hagedorn:2010th,Chen:2013wba}, that can very 
 well reproduce the CKM and PMNS matrices in agreement with the 
 experimental results, but where the simplicity in the matter content has been 
 left aside. Most of these models include a large number of flavons which are 
 required to accommodate correctly the mixings. A different approach consists 
 of extending the Higgs sector of the models. By itself, the SUSY $SU(5)$ 
 matter content provides the tools to accommodate masses and mixings via the 
 extension of the Higgs scalar sector, although gauge
 coupling unification may be compromised. This philosophy has been worked out 
 with success in 
 non-supersymmetric~\cite{Felix:2006pn,Mondragon:2007af,Canales:2012dr} and
 supersymmetric~\cite{Babu:2004tn,Kajiyama:2005rk,Kajiyama:2007pr,Babu:2009nn,Babu:2011mv} 
 scenarios where the concept of flavour has been extended to the Higgs sector.

 Therefore, we propose here a renormalizable SUSY $SU(5)$ model where
 the $Q_{6}$ group plays an important role in the flavour sector. The
 need to extend the scalar sector is evident in order to accommodate
 masses and mixings for quarks and leptons without breaking explicitly
 the flavour symmetry, so that three families of $H^{u}$ and $H^{d}$ $5$-plets 
 are introduced. At the same time, three RHN's are included by hand, and the type I see-saw 
 mechanism is invoked to get small masses for the active neutrinos. 
 At low energies when the first and second RHN masses are degenerate, a lower 
 bound for the reactor angle, $\theta_{13}^{\ell}$, is obtained. Namely,
 consistent results are obtained for the CKM matrix in the quark sector, 
 while in the leptonic sector, the values obtained for atmospheric and solar mixing angles are; 
 $\theta^{\ell^{th}}_{23} = \left( 46.18^{+0.66}_{-0.65} \right)^{\circ} $ and 
 $\theta^{\ell^{th}}_{12} = \left( 36.62 \pm 4.06 \right)^{\circ}$, which are consistent with 
 experimental data.
 However, the reactor mixing angle value 
 $\theta^{\ell^{th}}_{13} = \left(3.38^{+0.03}_{-0.02} \right)^{\circ}$, is not in
 good agreement with the central values of the global fits but it is
 still within the error bar of the different experiments,
 and fairly large in comparison to the tribimaximal scenario. This value 
 corresponds to a lower bound for the reactor mixing angle of the more general case 
 where the RHN's are not mass degenerate, similar to the case 
 of $S_3$ non-supersymmetric models, where relaxing the degeneracy condition 
 in the RHN masses gives the right value for 
 $\theta_{13}^{\ell}$~\cite{Canales:2012dr}. So, in order to enhance the theoretical 
 value of the reactor mixing angle, which must be in accordance with the current experimental data,    
 we consider that the two first RHN's, $N_{1}$ and $N_{2}$, have different masses in the Majorana 
 mass term. This assumption allows to get the following values for the normal [inverted] hierarchy: 
 $\theta_{12}^{ \ell^{th} } = \left( 34.71_{-0.98}^{+0.91} \right)^{\circ}$ 
 $\left[ \left( 34.73_{-1.11}^{+0.89} \right)^{\circ} \right]$, 
 $\theta_{23}^{ \ell^{th} } = \left( 45.83_{-3.98}^{+4.49} \right)^{\circ}$ 
 $\left[ \left( 48.57_{-2.76}^{+2.07} \right)^{\circ} \right]$, and 
 $\theta_{13}^{ \ell^{th} } = \left( 8.77_{-0.32}^{+0.40} \right)^{\circ}$ 
 $\left[ \left( 8.93_{-0.39}^{+0.33} \right)^{\circ} \right] $.

 The paper is organized as follows: In section 2, we build the extended SUSY 
 $SU(5)$ model which obeys the $Q_{6}$ flavour symmetry, in addition, we 
 explain the required matter content to get a unified scenario and we do 
 stress the strong assumptions that we will make in the model; at the same time, 
 the ansatz to increase the value of reactor mixing angle is discussed. The
 textures in the quark mass matrices, and therefore their consequences
 in the lepton sector, are analysed in section 3. Also, we describe how
 to diagonalize the mass matrices for each sector, 
 with particular attention to neutrino sector. 
 In section 4, we present and discuss the results
 about the mixing angles for each sector. Finally, we give conclusions on this 
 preliminary analysis of the model.

\section{SUSY $SU(5)\otimes Q_{6}$ model}
 We will consider the SUSY~$SU(5)$ model where three right-handed neutrinos have 
 been included in the matter content. In addition, the scalar sector has to be 
 necessarily extended as we will show. Thus, the matter content that will be
 used in the present model is displayed on Table~\ref{tabla1}, where the assignments
 under $Q_{6}$ flavour symmetry are shown.

\begin{table}[ht]
\begin{center} 
   \begin{tabular}{|c|c|c|} \hline \hline & 
{\small $SU(5)$} & {\small $Q_{6}$} \\ \hline 
{\footnotesize $\left( H^{d}_{1} , H^{d}_{2} \right)$} & 
{\footnotesize ${\bf\bar{5} }$} & 
{\footnotesize ${\bf 2 }_{1}$} \\ \hline
{\footnotesize $H^{d}_{3}$} &
{\footnotesize ${\bf \bar{5} }$} & 
{\footnotesize ${\bf 1 }_{+,2}$} \\ \hline
{\footnotesize $\left( H^{u}_{1} , H^{u}_{2} \right)$} &
{\footnotesize ${\bf5 }$} & 
{\footnotesize ${\bf 2 }_{1}$} \\ \hline
{\footnotesize $H^{u}_{3}$} & 
{\footnotesize ${\bf5 }$} & 
{\footnotesize ${\bf 1 }_{+,2}$} \\ \hline 
{\footnotesize $\left( F_{1} , F_{2} \right)$} & 
{\footnotesize ${\bf \bar{5} }$} & 
{\footnotesize ${\bf 2 }_{2}$} \\ \hline 
{\footnotesize $F_{3}$} & 
{\footnotesize ${\bf \bar{5} }$} & 
{\footnotesize ${\bf 1 }_{-,3}$} \\ \hline
{\footnotesize $\left( T_{1} , T_{2} \right)$} & 
{\footnotesize ${\bf 10}$} & 
{\footnotesize ${\bf 2 }_{2}$} \\ \hline
{\footnotesize $T_{3}$} & 
{\footnotesize ${\bf 10}$} & 
{\footnotesize ${\bf 1}_{-,3}$} \\ \hline
{\footnotesize $\left( N^{c}_{1} , N^{c}_{2} \right)$} & 
{\footnotesize ${\bf 1}$} & 
{\footnotesize ${\bf 2}_{2}$} \\ \hline
{\footnotesize $N^{c}_{3}$} & 
{\footnotesize ${\bf 1}$} & 
{\footnotesize ${\bf 1}_{-,1}$} \\ \hline 
{\footnotesize $Y_{B}$} & 
{\footnotesize ${\bf 1}$} & 
{\footnotesize ${\bf 1}_{+,2}$} \\ \hline
{\footnotesize $H_{\bar{45}}$} & 
{\footnotesize ${\bf \bar{45} }$} & 
{\footnotesize ${\bf 1}_{+,2}$}  \\ \hline
{\footnotesize $H_{45}$} & 
{\footnotesize ${\bf 45 }$} & 
{\footnotesize ${\bf 1 }_{+,2}$}  \\ \hline 
{\footnotesize $\Phi$} & 
{\footnotesize ${\bf 24 }$} & 
{\footnotesize ${\bf 1}_{+,0}$}  \\ \hline \hline  
   \end{tabular} 
   \caption{Matter content in the SUSY $SU(5) \otimes Q_{6}$ 
    model.}\label{tabla1}
  \end{center}
 \end{table}

 Let us comment on our notation and the matter content: $\Phi^{a}_{b}$
 stands for the ${\bf 24 }$ adjoint scalar representation which breaks
 ($\left\langle \Phi \right\rangle = \sigma \textrm{diag} \left( 1, 1,
   1, -3/2, -3/2 \right)$) the SUSY $SU(5)$ gauge group to the MSSM;
 there are three families of Higgs type $H^{u}_{i}$ and
 $H^{d}_{j}$. Here, we ought to stress a point. Because we
   have extended the Higgs sector, gauge coupling unification is not
   guaranteed, since the Renormalization Group Equations (RGE) depend
   strongly on the number of Higgs families \cite{Einhorn:1981sx,Astorga:1994gh}.
   This is a drawback in the present model, and it is the price we
   have to pay for going beyond the minimal SUSY $SU(5)$ model.  A
   proper analysis of the RGE evolution, including also the right
   handed-neutrinos will be left for a future work. On the other hand,
 we do need to include $H_{45}$ and ${H}_{\bar{45}}$ scalar
 representations so as to fix the incorrect relation ${\bf M}_{d} =
 {\bf M}^{T}_{e}$, although there are other ways to achieve it,
 see~\cite{Berezhiani:1998hg, Bajc:2002pg}. In addition, there is
 flavon, $Y_{B}$, which has already been used in supersymmetric
 models~\cite{Babu:2004tn, Kajiyama:2005rk, Kajiyama:2007pr,
   Babu:2009nn, Babu:2011mv} in order to have a flavour invariant
 Majorana mass matrix. Regarding the fermion sector, $N_{i}$ denotes
 the RHN, which is a singlet under the $SU(5)$ gauge group; $F_{i}$
 and $T_{j}$ stand for the ${\bf 5}$-plets and the ${\bf 10}$
 antisymmetric-plet, respectively. Here, $a, b,c$ are $SU(5)$ indices,
 and $i,j$ are family indices. More explicitly,
\begin{eqnarray}
  F_{i a} & = & \left(d^{c}, L\right); \quad 
  T^{ab}_{j} = \dfrac{1}{\sqrt{2}} \left( u^{c}, q, \ell^{c} \right); \quad 
  L = 
  \begin{pmatrix}
   \nu_{\ell} \\ 
   \ell
  \end{pmatrix}; \quad 
  q = 
  \begin{pmatrix}
   u \\ 
   d
  \end{pmatrix}, \\
  H^{u a} & = & 
  \begin{pmatrix}
   \texttt{H}^{u} \\ 
   \textbf{H}^{u}  
  \end{pmatrix}; \quad 
  H^{d}_{b}= \begin{pmatrix}
   \texttt{H}^{d} \\ 
   \textbf{H}^{d} 
  \end{pmatrix}; \quad  
  \textbf{H}^{u}= 
  \begin{pmatrix}
   h^{+ u} \\ 
   h^{0 u}
  \end{pmatrix}; \quad  
  \textbf{H}^{d}= 
  \begin{pmatrix}
   h^{0 d} \\ 
   h^{- d}
   \end{pmatrix} .\nonumber \label{eq1}
 \end{eqnarray}
 Here, $\texttt{H}^{u}$ and $\texttt{H}^{d}$ are the coloured triplet scalars 
 that mediate proton decay. For the time being, it will be assumed that these 
 are heavy enough to keep proton lifetime bounded and under control. In 
 addition, we have to point out that this subtle issue will be left aside, 
 since we are only interested in studying the masses and mixings when 
 implementing $Q_{6}$ as a flavour symmetry in this model. On the other 
 hand, $\textbf{H}^{d}$ and $\textbf{H}^{u}$ are the weak doublets of the MSSM 
 group. In consequence, we employ the following vacuum expectation values (vev's) to get 
 the fermion mass matrices.
 
 \begin{eqnarray}
  \langle \textbf{H}^{u }\rangle & = & 
  \begin{pmatrix}
   0 \\ 
   h^{0u}
  \end{pmatrix}, \qquad  
  \langle \textbf{H}^{d} \rangle = 
  \begin{pmatrix}
   h^{0 d} \\ 
   0
  \end{pmatrix},\qquad
  \left\langle H_{45} \right\rangle^{\alpha 5}_{\alpha} = v_{45},\qquad   
  \left\langle H_{45} \right\rangle^{4 5}_{4} = -3 v_{45} \nn\\
  \left\langle H_{\bar{45}} \right\rangle^{\alpha}_{\alpha 5} & = & 
  v_{\bar{45}},\qquad 
  \left \langle H_{\bar{45}} \right \rangle^{4}_{4 5} = -3 v_{\bar{45}},\qquad \left \langle Y_{B} \right 
  \rangle=v_{B}
  \qquad 
  \alpha,\beta=1,2,3. \label{eq2}
 \end{eqnarray}
 
 Having presented the assigned matter fields under the $Q_{6}$ flavour symmetry, 
 we will now introduce the superpotential, gauge invariant under the $Q_{6}$ 
 discrete group. All necessary details, on the multiplication rules of $Q_{6}$, are given in the appendix, where a  brief review of this dihedral 
 group is offered. The trilinear terms in the superpotential are given by
{\footnotesize
\begin{align}
W &=\sqrt{2} y^{d}_{1} \left( F_{1} T_{2} - F_{2} T_{1} \right) H^{d}_{3} 
+ \sqrt{2} y^{d}_{2} \left( F_{1} T_{3} H^{d}_{2} - F_{2} T_{3} H^{d}_{1}\right) 
+ \sqrt{2} y^{d}_{3} F_{3} \left( T_{1} H^{d}_{2} - T_{2} H^{d}_{1} \right) 
+ \sqrt{2} y^{d}_{4} F_{3} T_{3} H^{d}_{3}\nn\\& + \dfrac{ y^{u}_{1} }{4} 
\left( T_{1} T_{2} - T_{2} T_{1} \right) H^{u}_{3} 
+ \dfrac{ y^{u}_{2} }{4} \left( T_{1} T_{3} H^{u}_{2} 
- T_{2} T_{3} H^{u}_{1} \right)
+ \dfrac{ y^{u}_{3} }{4} T_{3} \left( T_{1} H^{u}_{2} -T_{2} H^{u}_{1} \right) + \dfrac{ y^{u}_{4} }{4} 
T_{3} T_{3} H^{u}_{3}\nn\\& + \sqrt{2} Y_{1} \left( F_{1} T_{2} -
F_{2} T_{1} \right) H_{\bar{45}} + \sqrt{2} Y_{2} F_{3} T_{3} H_{\bar{45}} 
+ \dfrac{ \tilde{Y}_{1}}{4} \left( T_{1} T_{2} - T_{2} T_{1} \right) H_{45}
+ \dfrac{ \tilde{Y}_{2}}{4} T_{3} T_{3} H_{45}\nn\\& + y^{n}_{1} \left( N^{c}_{1} 
F_{2} - N^{c}_{2} F_{1} \right) H^{u}_{3}
+ y^{n}_{2} \left( N^{c}_{1} F_{3} H^{u}_{2} - N^{c}_{2} F_{3} H^{u}_{1} 
\right) + y^{n}_{3} N^{c}_{3} \left( F_{1} H^{u}_{2} + F_{2} H^{u}_{1} 
\right) + M_{R_{1}} \left( N^{c}_{1} N^{c}_{1} + N^{c}_{2} 
N^{c}_{2} \right) \nonumber\\& 
+ y^{m}_{2} N^{c}_{3} Y_{B} N^{c}_{3}. \label{eqq3}
 \end{align}}
 It is important to remember 
 that SUSY must be broken via soft breaking terms, so these soft breaking 
 terms should be included in a complete study of the full scalar potential,
 but we do not include them in this preliminary analysis. The  scalar  
 superpotential is
{\footnotesize 
 \begin{align}
  W_{s} &= \mu_{\Phi} Tr \left( \Phi^{2} \right) 
   + \lambda_{\Phi} Tr \left( \Phi^{3} \right) 
   + m_{45} H_{\bar{45}} H_{45} 
   + \mu_{1} \left( H^{u}_{1} H^{d}_{2} - H^{u}_{2} H^{d}_{1} \right) 
   + \mu_{2} H^{u}_{3} H^{d}_{3}\nn\\&+ \lambda_{1} \left( H^{u}_{1} \Phi H^{d}_{2} 
   - H^{u}_{2} \Phi H^{d}_{1} \right) + \lambda_{2} H^{u}_{3} \Phi H^{d}_{3} 
   + a_{1} H_{\bar{45}} H_{45} \Phi 
   + a_{2} H_{\bar{45}} \Phi H^{u}_{3} + a_{3} H^{d}_{3} \Phi H_{45}. 
  \label{EQX}
 \end{align}}
 From the superpotential given in Eq.~(\ref{eqq3}), one must obtain the MSSM 
 effective superpotential that contains the Yukawa mass term after spontaneous 
 symmetry breaking via the vev's of the $\textbf{H}^{u}$ and $\textbf{H}^{d}$ 
 weak doublets scalar superfields. We will work in the following basis
 \begin{equation}\label{eq3}
  \mathcal{L} = \mathcal{L}^{q} + \mathcal{L}^{l}
 \end{equation}
 with
 \begin{equation}\label{masster}
  \begin{array}{l}
   \mathcal{L}^{q} = - \bar{d}_{i R} \left( {\bf M}_{d} \right)_{ij} d_{j L} 
    - \bar{u}_{i R} \left( {\bf M}_{u} \right)_{ij} u_{j L}  + h.c. \, , \\
   \mathcal{L}^{l} =  \bar{\ell}_{i R} \left( {\bf M}_{\ell} \right)_{ij} 
    \ell_{j L} - \bar{N}_{i R} \left( {\bf M}_{D} \right)_{ij} \nu_{j L}
    - \frac{1}{2} \bar{N}_{i R} \left( {\bf M}_{R} \right)_{ij} N^{c}_{j R} 
    + h.c.\, ,  
  \end{array}
 \end{equation} 
 In general, the up, down, and charged lepton mass matrices are given by 
 (see~\cite{Dorsner:2007fy}) 
{\scriptsize
\begin{align}\label{matrixfor}
 {\bf M}_{u} = 
   \dfrac{ \left( {\bf Y}^{u} + {\bf Y}^{u T} \right) }{2} 
    \langle \textbf{H}^{u} \rangle 
    - \left( \tilde{{\bf Y}}^{T} - \tilde{{\bf Y}} \right) v_{45}, \quad 
    {\bf M}_{d} = {\bf Y}^{d} \langle \textbf{H}^{d} \rangle 
    + 2 {\bf Y} v_{\bar{45}}, \quad 
   {\bf M}_{\ell} = {\bf Y}^{d T} \langle \textbf{H}^{d} \rangle 
    -6 {\bf Y}^{T} v_{\bar{45}} .
 \end{align}}
 In this particular model, from Eqs.~(\ref{eqq3}) and~(\ref{matrixfor}) we 
 have that the up, down and charged lepton mass matrices have respectively the 
 following structures:
{\scriptsize
\begin{align}\label{eq4}
  {\bf M}_{u} &=
  \begin{pmatrix}
   0 & -2\tilde{Y}_{1} v_{45} &  \bar{y}^{u} h^{0 u}_{2} \\ 
   2 \tilde{Y}_{1} v_{45} & 0 & - \bar{y}^{u} h^{0 u}_{1} \\ 
   \bar{y}^{u} h^{0 u}_{2} & - \bar{y}^{u}h^{0 u}_{1}  & y^{u}_{4} h^{0 u}_{3}
  \end{pmatrix},\,
  {\bf M}_{d} = 
 \begin{pmatrix}
   0 & y^{d}_{1} h^{0 d}_{3} + 2 Y_{1} v_{\bar{45}} & y^{d}_{2} h^{0 d}_{2} \\ 
   - y^{d}_{1} h^{0 d}_{3} - 2 Y_{1} v_{\bar{45}} & 0 & 
    - y^{d}_{2} h^{0 d}_{1} \\ 
   y^{d}_{3} h^{0 d}_{2} & - y^{d}_{3} h^{0 d}_{1} & y^{d}_{4} 
    h^{0 d}_{3} + 2Y_{2} v_{\bar{45}}
  \end{pmatrix}; \nonumber\\
  {\bf M}_{\ell} & = 
  \begin{pmatrix}
   0 & - \left( y^{d}_{1} h^{0 d}_{3} - 6 Y_{1} v_{\bar{45}} \right) & 
    y^{d}_{3} h^{0 d}_{2} \\ 
   y^{d}_{1} h^{0 d}_{3} - 6 Y_{1} v_{\bar{45}} & 0 & - y^{d}_{3} 
    h^{0 d}_{1} \\ 
   y^{d}_{2} h^{0 d}_{2} & - y^{d}_{2} h^{0 d}_{1} & y^{d}_{4} 
    h^{0 d}_{3} - 6 Y_{2} v_{\bar{45}}
  \end{pmatrix}.
\end{align}} where $\bar{y}^{u} \equiv \left( y^{u}_{2} + y^{u}_{3}
\right)/2$.  As can be seen, the ${\bf M}_{u}$ mass matrix turns out
almost symmetric due to the flavour structure. At the same time, we
were able to correct the wrong relationship between the down quarks
and the charged leptons; this was achieved including the
$H_{\bar{45}}$ scalar representation. From
  Eq.(\ref{eqq3}), we obtain the Dirac (see Eq.~(\ref{eq5})) and
  Majorana mass matrices, this latter is strictly given by ${\bf
    m}_{R} = \textrm{diag} \left( M_{R_{1}}, M_{R_{1}}, M_{R_{3}} =
    y^{m}_{2}v_{B} \right)$.  We would like to mention at this point
  the ansatz to enhance the $\theta_{13}^{l}$ value, and its subtle
  ingredient, which consists in assuming that two RHN's, $N_{1}$ and
  $N_{2}$, are not degenerated. Thus, the Majorana mass matrix is
  given in Eq. (\ref{eq5}), and two cases will be studied later:
  $a)$~RHN's degenerate, $b)$~RHN's non-degenerate.  
{\small
\begin{align}\label{eq5}
 {\bf M}_{D} = 
   \begin{pmatrix}
   0 & y^{n}_{1} h^{0 u}_{3} & y^{n}_{2} h^{0 u}_{2} \\ 
   -y^{n}_{1} h^{0 u}_{3}& 0 &-y^{n}_{2} h^{0 u}_{1} \\ 
   y^{n}_{3} h^{0 u}_{2} & y^{n}_{3} h^{0 u}_{1} & 0
  \end{pmatrix}
  \quad \textrm{and} \quad
  {\bf M}_{R} = 
  \begin{pmatrix}
   M_{R_{1}}& 0 & 0 \\ 
   0 & M_{R_{2}} & 0 \\ 
   0 &  & M_{R_{3}}
  \end{pmatrix}.
  \end{align}}
Therefore, after the type I see-saw mechanism, the 
 neutrino mass term is given by 
 \begin{equation}\label{eqextra}
  \mathcal{L}_{\nu} = -\bar{N}_{i R} \left( {\bf M}_{D} \right)_{ij} \nu_{j L} 
   - \frac{1}{2} \bar{N}_{i R} \left( {\bf M}_{R} \right)_{ij} N^{c}_{j R} 
   + h.c. 
  = - \frac{1}{2} \bar{\nu}^{C}_{L} {\bf M}_{\nu} \nu_{L} 
   - \frac{1}{2} \bar{N}_{i R} \left( {\bf M}_{R} \right)_{ij} N^{c}_{j R}.
 \end{equation} 
 where the ${\bf M}_{\nu} = {\bf M}^{T}_{D} {\bf M}_{R}^{-1} {\bf M}_{D}$ 
 effective neutrino mass matrix has the following structure:
{\small
\begin{align}\label{eq6}
  {\bf M}_{\nu} =
  \left( \begin{array}{ccc}\vspace{2mm}
  \frac{ \left( y^{n}_{1} h^{0 u}_{3} \right)^{2}}{ M_{R_{2}} } 
   + \frac{ \left( y^{n}_{3} h^{0 u}_{2} \right)^{2}}{ M_{R_{3}} } &
  \frac{ \left( y^{n}_{3} \right)^{2} h^{0 u}_{2} h^{0 u}_{1} }{ M_{R_{3}} } &
  \frac{ y^{n}_{1} h^{0 u}_{3} y^{n}_{2} h^{0 u}_{1} }{ M_{R_{2}} } 
   \\ \vspace{2mm}
  \frac{ \left( y^{n}_{3} \right)^{2} h^{0 u}_{2} h^{0 u}_{1} }{ M_{R_{3}} } & 
  \frac{ \left( y^{n}_{1} h^{0 u}_{3} \right)^{2}}{ M_{R_{1}} } 
   + \frac{ \left( y^{n}_{3} h^{0 u}_{1} \right)^{2}}{ M_{R_{3}} } & 
  \frac{ y^{n}_{1} h^{0 u}_{3} y^{n}_{2} h^{0 u}_{2} }{ M_{R_{1}} } 
   \\\vspace{2mm}
  \frac{ y^{n}_{1} h^{0 u}_{3} y^{n}_{2} h^{0 u}_{1} }{ M_{R_{2}} } & 
  \frac{ y^{n}_{1} h^{0 u}_{3} y^{n}_{2} h^{0 u}_{2} }{ M_{R_{1}} } & 
  \frac{ \left( y^{n}_{2} h^{0 u}_{2} \right)^{2}}{ M_{R_{1}} } 
   + \frac{ \left( y^{n}_{2} h^{0 u}_{1} \right)^{2}}{ M_{R_{2}} }
  \end{array}   \right).
 \end{align}}
 At first sight the mass matrix $M_{\nu}$ may seem very complicated to be 
 diagonalized analytically. Also, it has twelve real free parameters, since it is a 
 complex symmetric matrix.
 Thus, the number of parameters in the matrix $M_{\nu}$ is a very large. However, 
 it is well known that an effective way to reduce the number of 
 parameters in a mass matrix is to perform a similarity transformation, through 
 which it is possible to go to a basis  where the matrix will have some texture 
 zeros~\cite{Fritzsch:1999ee}. Then, in complete analogy with the work done on 
 the $S_{3}$-flavour 
 symmetry~\cite{Canales:2013ura,GonzalezCanales:2012kj,GonzalezCanales:2012za,Canales:2012dr} 
 and due to the form of ${\bf M}_{\nu}$, we can rotate the left-handed neutrino 
 field as follows: $ \nu_{L} = {\bf U}_{\nu} \tilde{\nu}_{L}$, where 
 ${\bf U}_{\nu} = {\bf \textit{u} }_{\theta} {\bf u}_{\nu}$ so that for a 
 normal [inverted] hierarchy in the neutrino masses one gets\footnote{Here we 
 are assuming a hierarchical ansatz, i.e., the particle masses are arranged 
 from lightest to heaviest, placing the last one in the (3,3) position of the 
 diagonal mass matrix.}:
 \begin{equation}
  {\bf \tilde{M} }_{\nu} = 
   \textrm{diag} \left( m_{\nu_{1[3]}}, m_{\nu_{2[1]}}, m_{\nu_{3[2]}} 
    \right) 
  = {\bf u}^{T}_{\nu}{\bf m}_{\nu} {\bf u}_{\nu},
 \end{equation}  
 where  
 \begin{equation}\label{ML:ec:13}
  {\bf m}_{\nu} = 
  {\bf \textit{u} }_{\theta}^{T} {\bf M}_{\nu} {\bf \textit{u} }_{\theta} 
  =
  \left( \begin{array}{ccc}
   b_{\nu} & a_{\nu} & c_{\nu} \\
   a_{\nu} & \mu_{0} & 0       \\
   c_{\nu} & 0       & d_{\nu}
  \end{array}   \right) 
  \quad \textrm{and} \quad 
  {\bf \textit{u} }_{\theta} =
  \left( \begin{array}{ccc}
   \cos \theta & 0 & -\sin \theta \\
   \sin \theta & 0 &  \cos \theta \\
   0           & 1 & 0
  \end{array}   \right),
 \end{equation}  
 with  
 {\small \begin{equation}\label{ML:ec:14}
  \begin{array}{l} \vspace{2mm}
  \tan \theta = 
   \frac{ M_{R_{2}} }{ M_{R_{1}} } \frac{ h^{0 u}_{2} }{ h^{0 u}_{1} }, \quad
   \mu_{0} = \frac{ \left( y_{2}^{n} \right)^{2} }{ M_{R_{1}} M_{R_{2}} } 
   \left[ \left( h^{0 u}_{2} \right)^{2} M_{R_{2}} + \left( h^{0 u}_{1} 
   \right)^{2} M_{R_{1}} \right] , \\ \vspace{2mm}
  a_{\nu} = 
   \frac{ \cos \theta y_{1}^{n} h^{0 u}_{3} y_{2}^{n} }{ M_{R_{2} } 
   M_{ R_{1} }^{2} h^{0 u}_{1} } \left( \left( h^{0 u}_{1} \right)^{2} 
   M_{ R_{1} }^{2} + \left( h^{0 u}_{2} \right)^{2} M_{R_{2}}^{2} 
   \right), \\ \vspace{2mm}
  b_{\nu} = 
   \frac{ \cos^{2} \theta}{ M_{R_{1}}^{2} } 
   \left[ \frac{ \left( y_{1}^{n} h^{0 u}_{3} \right)^{2} }{ M_{ R_{1} } 
   M_{ R_{2} } \left( h^{0 u}_{1} \right)^{2} } 
   \left( M_{R_{1}}^{3} \left( h^{0 u}_{1} \right)^{2} + M_{R_{2}}^{3} 
   \left( h^{0 u}_{2} \right)^{2} \right) 
   + \frac{ \left( y_{3}^{n} h^{0 u}_{2} \right)^{2} }{ M_{R_{3}} } 
   \left( M_{R_{1}} + M_{R_{2}} \right)^{2} \right], \\ \vspace{2mm}
  c_{\nu} =  
   \frac{ h^{0 u}_{2} \cos^{2} \theta }{ h^{0 u}_{1} M_{R_{1}}^{2} } 
   \left[ \left( y_{1}^{n} h^{0 u}_{3} \right)^{2} \left( M_{R_{2}} -
   M_{R_{1}} \right) 
   + \frac{ \left( y_{3}^{n} \right)^{2} }{ M_{R_{3}} } \left( M_{R_{1}} 
   + M_{R_{2}} \right) \left( \left( h^{0 u}_{1} \right)^{2} M_{R_{1}} 
   - \left( h^{0 u}_{2} \right)^{2} M_{R_{2}} \right)  \right],
   \; \textrm{and} \; \\ \vspace{2mm} 
  d_{\nu} = 
   \frac{ \cos^{2} \theta }{ \left( M_{R_{1}} h^{0 u}_{1} \right)^{2} } 
   \left[ \left( y_{1}^{n} h^{0 u}_{3} \right)^{2} \left( M_{R_{2}} 
   \left( h^{0 u}_{2}  \right)^{2}  + M_{R_{1}} \left( h^{0 u}_{1}  
   \right)^{2} \right) + \frac{ \left( y_{3}^{n} \right)^{2} }{ M_{R_{3}} } 
   \left( M_{R_{1}} \left( h^{0 u}_{1} \right)^{2} - M_{R_{2}} 
   \left( h^{0 u}_{2}  \right)^{2} \right)^{2} \right].
  \end{array}  
 \end{equation} } 
 The mass matrix ${\bf m}_{\nu}$, Eq.~(\ref{ML:ec:13}), has one texture zero, 
 so it has ten real effective free parameters. But, as we will show later on, when we relate 
 ${\bf m}_{\nu}$ with a matrix with two texture zeroes of class
 I~\cite{Canales:2013ura,GonzalezCanales:2012kj,GonzalezCanales:2012za,Canales:2012dr,Canales:2013cga}, 
 the number of real effective free parameters is reduced to only four. 
 
\section{Masses and mixings in the NNI scenario}

There are two ways to obtain the Fritzsch and NNI
  textures in the up, down quark and charged lepton sector,
  respectively. The first scenario consists in taking the condition
  $h^{0 u}_{2}=0=h^{0 d}_{2}$ on the vev's, and the second one
  consists in assuming $h^{0 u}_{2}=h^{0 u}_{1}\equiv h^{0 u}$ and
  $h^{0 d}_{2}=h^{0 d}_{1}\equiv h^{0 d}$. Without loss of generality,
  in the next subsection we will describe the second way.

\subsection{Quark and lepton masses}
 First, if we assume that $h^{0 u}_{2}=h^{0 u}_{1}\equiv h^{0 u}$ and 
 $h^{0 d}_{2}=h^{0 d}_{1}\equiv h^{0 d}$, we obtain the following mass  
 matrices:
{\small
\begin{align}
  {\bf M}_{j} =
  \begin{pmatrix}
   0 & \pm A_{j} & B_{j} \\ 
   \mp A_{j} & 0 & -B_{j} \\ 
   C_{j} & -C_{j} & D_{j}
  \end{pmatrix}, 
  \quad \textrm{and} \quad
  {\bf M}_{u} = 
  \begin{pmatrix}
   0 & -A_{u} & B_{u} \\ 
   A_{u} & 0 & -B_{u} \\ 
   B_{u} & -B_{u} & D_{u}
  \end{pmatrix}.\label{eq9}
 \end{align}}
 Here, the subindex $j$ denotes the charged lepton and quark-down mass matrices, 
 namely $j= \ell, d$. While the upper (lower) sign corresponds to the ${\bf M}_{d}$
 (${\bf M}_{\ell}$) mass matrix. In addition, the explicit form of the matrix 
 elements is given in
 Eq.~(\ref{eq4}). The ${\bf M}_{j}$ and ${\bf M}_{u}$ mass matrices 
 contain implicitly the $NNI$ and $Fritzsch$ textures respectively which 
 appear explicitly as follows: the above mass matrices are diagonalized by 
 unitary matrices, ${\bf U}_{k (R,L)}$ with the subindex $k=d, \ell, u$. According to 
 Eq.~(\ref{masster}) one obtains $\tilde{{\bf M}}_{ k } = 
 {\bf U}^{\dagger}_{k R} \hat{ {\bf M} }_{k} {\bf U}_{k L}$, in general. Here, 
 $\tilde{{\bf M}}_{k} = \textrm{diag} 
 \left( \tilde{m}_{k_{1}}, \tilde{m}_{k_{2}}, 1 \right)$. For simplicity, we 
 have normalized the above expressions so that 
 $\tilde{m}_{k_{1}} = m_{k_{1}}/m_{k_{3}}$,
 $\tilde{m}_{k_{2}} = m_{k_{2}}/m_{k_{3}}$ and 
 $\hat{{\bf M}}_{k} = {\bf M}_{k}/m_{k_{3}}$ are dimensionless parameters.
 Then, taking the unitary matrices as ${\bf U}_{k (R, L)} = {\bf U}_{\pi/4} {\bf u}_{k (R, L)}$, one 
 can get easily $\tilde{{\bf M}}_{k}={\bf u}^{\dagger}_{k R} {\bf m}_{k} 
 {\bf u}_{k L}$ where  
{\small
 \begin{align}\label{eq10} 
  {\bf m}_{k} = {\bf U}^{T}_{\pi/4} \hat{{\bf M}}_{k} {\bf U}_{\pi/4} = 
\begin{pmatrix}
   0 & \pm \tilde{A}_{k} & 0 \\ 
   \mp \tilde{A}_{k} & 0 & -\sqrt{2}\tilde{B}_{k} \\ 
   0 & -\sqrt{2}\tilde{C}_{k} & \tilde{D}_{k}
  \end{pmatrix},
  \; \textrm{and}\; 
  {\bf U}_{\pi/4} = 
  \begin{pmatrix}\vspace{2mm}
   \frac{1}{ \sqrt{2} } & - \frac{1}{ \sqrt{2} } & 0 \\ \vspace{2mm}
   \frac{1}{ \sqrt{2} } &   \frac{1}{ \sqrt{2} } & 0 \\ 
   0 & 0 & 1
  \end{pmatrix}.
 \end{align}}

We should point out that the $\sqrt{2}$ factor will be absorbed in the
$\tilde{B}$ and $\tilde{C}$ dimensionless free parameters for the
quark and lepton sectors, respectively. In addition, notice that for
the ${\bf m}_{u}$ mass matrix given in Eq.~(\ref{eq10}),
$\tilde{C}_{u}=\tilde{B}_{u}$, according to
Eq. (\ref{eq9}). In the neutrino sector, the
  degeneracy, $h^{0 u}_{2}=h^{0 u}_{1}\equiv h^{0 u}$, in the
  effective mass matrix given in Eq.~(\ref{ML:ec:13}) reduces
  substantially the free parameters when two RHN's are degenerated;
  this is not true for the non-degenerate case where this assumption
  does not modify the functional structure of the effective neutrino
  mass matrix given in Eq.~(\ref{ML:ec:13}), because this degeneration
  only simplifies a little the form of entries of matrix ${\bf
    m}_{\nu}$, for more details see the appendix~\ref{ApenC}.
 
 Let us add a comment on the degeneracy on the vacuum expectation values. As 
 we have remarked, the conditions $h^{0 u}_{2}=h^{0 u}_{1}\equiv h^{0 u}$ and 
 $h^{0 d}_{2}=h^{0 d}_{1}\equiv h^{0 d}$ have been assumed so far, it is not 
 clear yet that these relations will arise in a natural way upon minimization 
 of the scalar potential.  We expect that it will be the case and the study of 
 the full scalar potential may be done along the lines  given 
 in~\cite{Babu:2004tn, Babu:2009nn}.

\subsection{Quark and lepton mixings}
 We will describe briefly how to diagonalize the mass matrices, 
 ${\bf m}_{j}$ and ${\bf m}_{u}$ $(j=d,\ell)$, respectively. Let us first start with 
 the down quark and charged lepton mass matrices which have the NNI 
 textures, we will not enter in great detail since these kind of matrices have 
 been well studied in \cite{Harayama:1996am,Harayama:1996jr}. The above 
 mentioned description is applied to the ${\bf m}_{u}$ mass matrix where the 
 Fritzsch texture~\cite{Fritzsch:1977vd, Fritzsch:1977za,Fritzsch:2011cu} is
 present. For a pedagogical method to diagonalize these mass matrices 
 see~\cite{Felix:2006pn}.

 Going back to the expression 
 $\tilde{{\bf M}}_{k}={\bf u}^{\dagger}_{k R} {\bf m}_{k} {\bf u}_{k L}$, we 
 are interested in obtaining the ${\bf u}_{k L}$ left-handed matrices that 
 appear in the $CKM$ matrix, and for this we must build the bilineal form: 
 ${\bf \tilde{M}}^{\dagger}_{k} {\bf \tilde{M}}_{k} = {\bf u}^{\dagger}_{k L} 
 {\bf m}^{\dagger}_{k} {\bf m }_{k} {\bf u}_{k L}$. From this relation, we
 can factorize the CP phases that come from ${\bf m }^{\dagger}_{k}
 {\bf m }_{k} = {\bf Q}_{k} \left( {\bf m}^{\dagger}_{k} {\bf m}_{k} \right) 
 {\bf Q}^{\dagger}_{k}$, see~\cite{Canales:2011ug}, such that
{\small
 \begin{align}
   {\bf Q}_{k} = \textrm{diag} \left( 1, e^{-i \eta_{k_{2}}}, e^{-i \eta_{k_{3}}} \right) & 
   \textrm{  and} &
  ( {\bf m}^{\dagger}_{k} {\bf m}_{k} ) =
 \begin{pmatrix}
   \vert \tilde{A}_{k} \vert^2  & 0 & \vert \tilde{A}_{k} \vert 
    \vert \tilde{B}_{k} \vert \\ 
   0 & \vert \tilde{A}_{k} \vert^2 + \vert \tilde{C}_{k} \vert^2 & 
    \vert \tilde{C}_{k} \vert \vert \tilde{D}_{k} \vert \\ 
   \vert \tilde{A}_{k} \vert \vert \tilde{B}_{k} \vert & 
   \vert \tilde{C}_{k} \vert \vert \tilde{D}_{k} \vert & 
   \vert \tilde{B}_{k} \vert^2 +\vert \tilde{D}_{k} \vert^2
  \end{pmatrix}, \label{eq13}
 \end{align}}
 we should keep in mind that the 
 $\vert \tilde{A}_{k} \vert$, $\vert \tilde{B}_{k} \vert$, 
 $\vert \tilde{C}_{k} \vert$ and the $\vert \tilde{D}_{k} \vert$ free 
 parameters are real and dimensionless, and that for the up-quark mass matrix we
 have that $\vert \tilde{B}_{u} \vert = \vert \tilde{C}_{u} \vert$.

 Having factorized out the phases associated with CP violation in the bilineal 
 form, we choose  ${\bf u}_{k L} = {\bf Q}_{k} {\bf O}_{k L}$, where 
 ${\bf O}_{k L}$ is the orthogonal matrix that diagonalizes to the 
 $\left( {\bf m}^{\dagger}_{k} {\bf m }_{k} \right)$ matrix.
 The matrix ${\bf O}_{k L}$ is given by
 \begin{equation}\label{eq14}
  {\bf O}_{k L}= \left( \mid f_{1} \rangle, \mid f_{2} \rangle , 
   \mid f_{3} \rangle \right), 
 \end{equation}
 where the three eigenvectors have the following form
{\footnotesize
 \begin{align}\label{eq14.4}
  \mid f_{i} \rangle = N_{f_{i}} 
  \begin{pmatrix}
   \left( \tilde{m}^{2}_{ k_{i} }- \vert \tilde{A}_{k} \vert^2 
    - \vert \tilde{C}_{k} \vert^2 \right) \vert \tilde{A}_{k} \vert 
    \vert \tilde{B}_{k} \vert \\ 
   \left( \tilde{m}^{2}_{k_{i}} - \vert \tilde{A}_{k} \vert^2 \right) 
    \vert \tilde{C}_{k} \vert \vert \tilde{D}_{k} \vert  \\ 
   \left( \tilde{m}^{2}_{k_{i}} - \vert \tilde{A}_{k} \vert^2 \right)
   \left( \tilde{m}^{2}_{k_{i}} - \vert \tilde{A}_{k} \vert^2 
   - \vert \tilde{C}_{k} \vert^2 \right)
  \end{pmatrix}, \,
 \mid f_{3} \rangle = N_{f_{3}}
  \begin{pmatrix}
   \left( 1-\vert \tilde{A}_{k} \vert^2-\vert \tilde{C}_{k} \vert^2 \right) 
   \vert \tilde{A}_{k} \vert \vert \tilde{B}_{k} \vert \\ 
   \left( 1 -\vert \tilde{A}_{k} \vert^2 \right) \vert \tilde{C}_{k} \vert 
   \vert \tilde{D}_{k} \vert  \\ 
   \left( 1 - \vert \tilde{A}_{k} \vert^2 \right) 
   \left( 1 - \vert \tilde{A}_{k} \vert^2 - \vert \tilde{C}_{k} \vert^2 
   \right)
  \end{pmatrix}.
 \end{align}}
 Here, $N_{f_{i}}$ ($i=1,2$) and $N_{f_{3}}$ stand for the
 normalization factors whose definition must be read directly from the
 above expression. 
 
 On the other hand, for the down quarks and charged leptons, 
 three free parameters can be fixed in terms of the physical masses and 
 $\vert \tilde{D}_{j} \vert \equiv y_{j}$, with $j = d, \ell$~\cite{Harayama:1996jr,Harayama:1996am}. Explicitly, these are given by
{\small
 \begin{align}\label{eq15}
  \vert \tilde{A}_{j} \vert = \frac{ q_{j} }{ y_{j} }, \, 
  \vert \tilde{B}_{j} \vert = \sqrt{ \frac{ 1 + P_{j} - y^{4}_{j} - R_{j} }{2} 
   - \left( \frac{ q_{j} }{ y_{j} } \right)^{2} }, \,
  \vert \tilde{C}_{j} \vert = \sqrt{ \frac{ 1 + P_{j} - y^{4}_{j} + R_{j} }{2} 
   - \left( \frac{ q_{j} }{ y_{j} } \right)^{2}} ,
 \end{align}}
 where
{\small
 \begin{align}\label{eq16}
   P_{j} = \tilde{m}_{ j_{1} }^{2} + \tilde{m}_{ j_{2} }^{2}, \, 
   q_{j} = \sqrt[4]{ \tilde{m}_{ j_{1} }^{2} \tilde{m}_{ j_{2} }^{2} },\, 
   R_{j} = \sqrt{\left( 1 + P_{j} -   y^{4}_{j} \right)^{2} 
    - 4 \left( P_{j} + q^{4}_{j} \right) + 8 q^{2}_{j} y^{2}_{j} } .
 \end{align}}
 In the above expressions there is only one free parameter which is $y_{j}$. This
 parameter should be tuned in order to get reliable mixing matrices
 as we will see later. So far, we have found the ${\bf U}_{j,L}$
 left-handed matrices that diagonalize the ${\bf M}_{j}$ mass
 matrices which have the NNI textures. Let us now focus on 
 ${\bf m}_{u}$. Going back to Eq.~(\ref{eq13}), we must remember that
 $\vert \tilde{B}_{u} \vert = \vert \tilde{C}_{u} \vert$, then one can
 determine the three free parameters in terms of the physical masses. 
 Explicitly, we obtain 
{\small 
 \begin{align} \label{ortou} 
  \vert \tilde{A}_{u} \vert = 
   \sqrt{ \dfrac{\tilde{m}_{u} \tilde{m}_{c} }{ 1 - \tilde{m}_{c} + 
   \tilde{m}_{u} } },\quad 
  \vert \tilde{B}_{u} \vert = \sqrt{ \dfrac{ ( 1 - \tilde{m}_{c} )
   ( 1 + \tilde{m}_{u} )( \tilde{m}_{c} - \tilde{m}_{u} ) }
   { 1 - \tilde{m}_{c} + \tilde{m}_{u} } } ,\quad
  \vert \tilde{D}_{u} \vert = 1 - \tilde{m}_{c} + \tilde{m}_{u}.  
 \end{align}}
 
 Following the same procedure, the ${\bf O}_{u L}$ orthogonal matrix
 that diagonalizes $({\bf m}^{\dagger}_{u} {\bf m}_{u})$ is fixed in
 terms of above parameters. Thus, using the expression given in 
 Eq.~(\ref{eq14.4}), we get 
 \begin{align} \label{orthouu} 
 {\bf O}_{u L}=
 \begin{pmatrix}\vspace{2mm}
  - \sqrt{ \frac{ \tilde{m}_{c} ( 1 - \tilde{m}_{c} ) }{ 
    ( 1 - \tilde{m}_{u} ) ( \tilde{m}_{c} + \tilde{m}_{u} ) {\cal G}_{u} } } &  
  -\sqrt{ \frac{ \tilde{m}_{u} ( 1 + \tilde{m}_{u} ) }{ 
    ( 1 + \tilde{m}_{c} ) ( \tilde{m}_{c} + \tilde{m}_{u} ) {\cal G}_{u} } } & 
  \sqrt{ \frac{ \tilde{m}_{u} \tilde{m}_{c} ( \tilde{m}_{c} - \tilde{m}_{u}) }{ 
    ( 1 - \tilde{m}_{u} )( 1 + \tilde{m}_{c} ) {\cal G}_{u} } } \\ \vspace{2mm}
  -\sqrt{ \frac{ \tilde{m}_{u} ( 1 - \tilde{m}_{c} ) }{ 
    ( 1 - \tilde{m}_{u} ) ( \tilde{m}_{c} + \tilde{m}_{u} ) } } & 
  \sqrt{ \frac{ \tilde{m}_{c} ( 1 + \tilde{m}_{u} ) }{ 
    ( 1 + \tilde{m}_{c} ) ( \tilde{m}_{c} + \tilde{m}_{u} ) } }  & 
  \sqrt{ \frac{ ( \tilde{m}_{c} - \tilde{m}_{u} ) }{ 
    ( 1 - \tilde{m}_{u} )  ( 1 + \tilde{m}_{c} ) } } \\  \vspace{2mm}
  \sqrt{ \frac{ \tilde{m}_{u} ( 1 + \tilde{m}_{u} )( \tilde{m}_{c} - \tilde{m}_{u} ) }{ 
    ( 1 - \tilde{m}_{u} )( \tilde{m}_{c} + \tilde{m}_{u} ) {\cal G}_{u} } }  & 
  - \sqrt{ \frac{ \tilde{m}_{c} ( 1 - \tilde{m}_{c} ) ( \tilde{m}_{c} - \tilde{m}_{u} ) }{ 
    ( 1 + \tilde{m}_{c} ) ( \tilde{m}_{c} + \tilde{m}_{u} ) {\cal G}_{u} } } & 
  \sqrt{ \frac{ ( 1 + \tilde{m}_{u} )( 1 - \tilde{m}_{c} ) }{ 
    ( 1 - \tilde{m}_{u} )( 1 + \tilde{m}_{c} ) {\cal G}_{u} } } 
 \end{pmatrix} , 
 \end{align}
 where $ {\cal G}_{u} \equiv ( 1 - \tilde{m}_{c} + \tilde{m}_{u})$. From 
 expressions in Eq.~(\ref{ortou}) we get that real orthogonal matrix ${\bf O}_{u L}$ 
 does not has free parameters, since this only depends of the up-quark mass ratios. 
 
 Therefore,  
 the full left-handed unitary matrices that diagonalize the charged lepton, 
 down- and up-quark mass matrices are given by   
  \begin{equation}
  {\bf U}_{k L} = {\bf U}_{\pi/4} {\bf u}_{k L} = {\bf U}_{\pi/4} {\bf Q}_{k} {\bf O}_{k L},
   \qquad k = u, d, \ell,  
 \end{equation}

 Thus, the CKM mixing matrix may be completely determined and given by 
 \begin{equation}
  {\bf V}_{CKM}= {\bf U}^{\dagger}_{u L} {\bf U}_{d L} 
   = \left( {\bf U}_{\pi/4} {\bf Q}_{u} {\bf O}_{u L} \right)^{\dagger} 
   {\bf U}_{\pi/4} {\bf Q}_{d} {\bf O}_{d L}
   = {\bf O}^{T}_{u L} {\bf Q}_{q} {\bf O}_{d L},
 \end{equation}
 where we have defined 
 ${\bf Q}_{q} \equiv {\bf Q}^{\dagger}_{u} {\bf Q}_{d} =
 \textrm{diag} \left( 1, \, e^{ i\alpha}, \, e^{i \beta } \right)$ 
 with the phases factors 
 $\alpha = \eta_{u_{2}} - \eta_{d_{2}}$ and $\beta = \eta_{u_{3}} - \eta_{d_{3}}$, which 
 come from the quark mass matrices. These two phases can be related with the unique phase 
 $\delta_{KM}$ of the angle-phase parametrization used in the PDG~\cite{Nakamura:2010zzi}  
 by means of expression: 
 $\sin \delta_{KM} = {\cal J}_{q} \left( 1 - \left| V_{ub} \right|^{2}  \right)/
   \left| V_{ud} \right| \left| V_{tb} \right| \left| V_{us} \right| \left| V_{cb} \right| 
   \left| V_{ub} \right|$ 
 where ${\cal J}_{q}$ is the Jarlskog invariant. Also, the rotation matrix ${\bf U}_{\pi/4}$ 
 is unobservable in the quark flavour mixings. In this
 way, the quark mixing matrix 
 $V_{CKM}$ has only three free parameters which are $y_{d}$, $\alpha$ and $\beta$, since the
 quark mass ratios are not treated as free parameters because we allow their values to vary 
 within the experimental measurements region reported by PDG~\cite{Nakamura:2010zzi}.    
 In addition the CKM matrix can be obtained analytically or numerically, however, 
 we are now just interested in getting a numerical expression for
 it which will be done in the next section.

 Now, in the leptonic sector the flavour mixing matrix PMNS is  defined 
 as~\cite{Fritzsch:1999ee}
 \begin{equation}\label{eq:PMNS:26}
  {\bf V}_{PMNS} = {\bf U}^{\dagger}_{\ell L} {\bf U}_{\nu} {\bf K}.
 \end{equation}
 Here, the unitary matrix of charge leptons ${\bf U}_{\ell L}$, is written as; 
 ${\bf U}_{\ell L} = {\bf U}_{\pi/4} {\bf Q}_{\ell} {\bf O}_{\ell L}$.
 The explicit form of unitary matrix of neutrinos, ${\bf U}_{\nu}$, will be 
 obtained in next sections. We should point out that we will neglect the 
 ${\bf K}$ Majorana CP phases, which are unobservable in the magnitudes of 
 entries of the leptonic mixing matrix.

 Before diagonalizing the effective neutrino mass matrix, let us show the set 
 of neutrino observables which is considered along the analytic and numerical
 analysis~\cite{Schwetz:2011qt,Adamson:2011qu}. This is given below
 \begin{equation}\label{data}
  \begin{array}{l}
   \Delta m^{2}_{\odot} \left( 10^{-5} \textrm{eV}^{2} \right) = 
     m^{2}_{ \nu_{2} } - m^{2}_{ \nu_{1} } =  7.59^{+0.20}_{-0.18} \\\\
   \Delta m^{2}_{\textrm{ATM}}  \left( 10^{-3} \textrm{eV}^{2} \right) = 
    \left| m^{2}_{ \nu_{3} } - m^{2}_{ \nu_{1} } \right|
    = 2.50^{+0.09}_{-0.16} \left[ -2.40^{+0.08}_{-0.09} \right] \\\\
  \sin^{2}\theta_{12}^{\ell^{ex}} = 0.312^{+0.017}_{-0.015} \\\\
  \sin^{2}\theta_{23}^{\ell^{ex}} = 0.52^{+0.06}_{-0.07} \left[ 0.52 \pm 0.06 \right] \\\\
  \sin^{2}\theta_{13}^{\ell^{ex}} = 0.013^{+0.007}_{-0.005} 
   \left[ 0.016^{+0.008}_{-0.006} \right] \\\\
  \sin^{2}2\theta_{13}^{\ell^{ex}} = 0.076 \pm 0.068 ~~~\left( \textrm{MINOS} \right). 
  \end{array}
 \end{equation}
 Here, the data appearing in squad parentheses stand for the inverted case. Leaving 
 aside the experimental results, we focus on diagonalize  the neutrino mass 
 matrix ${\bf m}_{\nu}$. For this purpose in this paper we will consider two 
 cases: $a)$~the first two masses of the RHN's are degenerate. 
 $b)$~the RHN's masses are not degenerate. 
 
 In the first case we can only obtain a lower bound for the value of the reactor 
 angle~\cite{Canales:2012dr}. But it is important to consider the case when 
 $M_{R_{1}} = M_{R_{2}}$, because it gives us an idea of which are the allowed 
 values for the free parameters present in the leptonic mixing matrix. 
    
\subsubsection{The masses of the right-handed neutrinos with degeneration} 
 From the expressions in eqs.~(\ref{Ap:ec:13}) and~(\ref{Ap:ec:14}), given in the 
 appendix~\ref{ApenC}, it is very easy to see  that when the masses of the first two 
 RHN's are degenerate, $M_{R_{1}} = M_{R_{2}}$, the effective 
 neutrino mass matrix ${\bf m}_{\nu}$ is reduced to a block matrix, as shown 
 below: 
{\small
\begin{align} \label{eq18}
  {\bf m}_{\nu} = 
  {\bf \textit{u} }^{T}_{\theta=\pi/4} {\bf M}_{\nu} 
  {\bf \textit{u}}_{\theta=\pi/4} =
  \begin{pmatrix}
   A^{2}_{\nu} + 2 B^{2}_{\nu} &  \sqrt{2}A_{\nu} C_{\nu} & 0 \\ 
   \sqrt{2} A_{\nu} C_{\nu} & 2 C^{2}_{\nu} & 0 \\ 
   0 & 0 & A^{2}_{\nu}
  \end{pmatrix} 
  \quad \textrm{and} \quad 
  {\bf \textit{u} }_{\theta=\pi/4} =
  \begin{pmatrix}
   \frac{1}{\sqrt{2}} & 0 & -\frac{1}{\sqrt{2}} \\ 
   \frac{1}{\sqrt{2}} & 0 & \frac{1}{\sqrt{2}} \\ 
   0 & 1 & 0
  \end{pmatrix}. 
 \end{align}}
 The ${\bf m}_{\nu}$ block matrix can be easily diagonalized. First, let us 
 factorize the CP phases of ${\bf m}_{\nu}$~\cite{Canales:2011ug}. So that, 
 ${\bf m}_{\nu} = {\bf P}_{\nu} {\bf \hat{m} }_{\nu} {\bf P}_{\nu}$, where 
 ${\bf P}_{\nu}$ and ${\bf \hat{m}}_{\nu}$ are given in Eq.~(\ref{eq19}). We 
 can associate immediately $\vert A_{\nu} \vert^2 =m_{\nu_{3}}$. Thus, we
 just have to diagonalize the left upper block of ${\bf m}_{\nu}$ 
{\small 
 \begin{align}\label{eq19}
   {\bf P}_{\nu} = 
  \begin{pmatrix}
   e^{i\eta_{\nu_{1}}} & 0 & 0 \\ 
   0 & e^{i\eta_{\nu_{2}}} & 0 \\ 
   0 & 0 & e^{i\eta_{\nu_{1}}}
  \end{pmatrix}
  \, \textrm{and} \,
  {\bf \hat{m} }_{\nu}= 
  \begin{pmatrix}
   \vert A_{\nu} \vert^2 + 2 \vert B_{\nu} \vert^2 & 
   \sqrt{2} \vert A_{\nu} \vert \vert C_{\nu} \vert & 0 \\ 
   \sqrt{2} \vert A_{\nu} \vert \vert C_{\nu} \vert & 
   2 \vert C_{\nu} \vert^2 & 0 \\
   0 & 0 & \vert A_{\nu}\vert^2
  \end{pmatrix}.
 \end{align}}

 A necessary condition to factorize the phases in the above way is that the 
 $A^{2}_{\nu}$ and $B^{2}_{\nu}$ phases must be aligned, although they may be 
 different in magnitude. On the other hand, we appropriately choose 
 ${\bf u}_{\nu} = {\bf P}^{\dagger}_{\nu} {\bf O}_{\nu}$. Here, 
 ${\bf O}_{\nu}$ is a real orthogonal matrix that diagonalizes the 
 ${\bf \hat{m}}_{\nu}$ matrix. Using the left upper block of ${\bf m}_{\nu}$, 
 we fix the $\vert B_{\nu} \vert^2$ and $\vert C_{\nu}\vert^2$ free parameters 
 through the following equations
 \begin{eqnarray}
  m_{\nu_{2}} + m_{\nu_{1}} = \vert A_{\nu} \vert^2+ 2 \vert B_{\nu} \vert^2 
  + 2 \vert C_{\nu} \vert^2 & \textrm{and} & m_{\nu_{2}} m_{\nu_{1}} = 4 
  \vert B_{\nu} \vert^2 \vert C_{\nu} \vert^2. \label{eq21}
 \end{eqnarray}
 Solving for the rest of the free parameters we find that
 \begin{eqnarray}
  \vert B_{\nu} \vert^2_{\mp} = \frac{1}{4} \left( m_{\nu_{2}} + m_{\nu_{1}}
  - m_{\nu_{3}} \mp R_{\nu} \right)  & \textrm{and} &
  \vert C_{\nu} \vert^2_{\pm} = \frac{1}{4} \left( m_{\nu_{2}} + m_{\nu_{1}}
  - m_{\nu_{3}} \pm R_{\nu} \right),\label{eq21ex}
 \end{eqnarray}
 where 
 \begin{equation}
  R_{\nu} \equiv \sqrt{ \left( m_{\nu_{2}} + m_{\nu_{1}} 
  - m_{\nu_{3}} \right)^2 - 4 m_{\nu_{2}} m_{\nu_{1}}}.
 \end{equation}
 As we can observe, there are two solutions for $\vert B_{\nu} \vert^2$ and 
 $\vert C_{\nu} \vert^2$, respectively. However, following a straightforward 
 analysis, it is clear that one solution is  discarded by demanding that two 
 free parameters (for the normal and for the  inverted hierarchy) should be 
 real and positive definite since they come from a real symmetric matrix. As a result of this, we realize that the normal spectrum is ruled out. For 
 the inverted case ($m_{\nu_{2}}>m_{\nu_{1}}>m_{\nu_{3}}$),  
 $\vert B_{\nu} \vert^2_{-}$ and $\vert C_{\nu} \vert^2_{+}$ turn out being 
 real and positive, if and only if, the $m_{\nu_{3}}$ lightest neutrino mass 
 is very small. Actually, from the definition of $R_{\nu}$ we obtain the 
 following sum rule 
 \beq
  m_{\nu_{3}} \leq ( \sqrt{m_{\nu_{2}}} - \sqrt{m_{\nu_{1}}})^2, \label{eqsum}
 \eeq
 where the equality in the above expression means an upper bound for the 
 $m_{\nu_{3}}$ lightest mass. Having fixed $\vert B_{\nu} \vert^2_{-}$ and 
 $\vert C_{\nu} \vert^2_{+}$ in terms of the physical neutrino masses, the 
 ${\bf O}_{\nu}$ matrix is well determined by them. Explicitly, 
 ${\bf O}_{\nu}$ is given by
{\small
 \begin{align}\label{eq21exx}
  {\bf O}_{\nu} = 
 \begin{pmatrix}
   \sqrt{\dfrac{m_{\nu_{3}}\left(m_{\nu_{2}}+ m_{\nu_{1}}-m_{\nu_{3}} 
   + R_{\nu} \right) }{ \left( m_{\nu_{2}} - m_{\nu_{1}} \right) 
   \left( m_{\nu_{2}} - m_{\nu_{1}} + m_{\nu_{3}} - R_{\nu} \right) }} & 
   \sqrt{ \dfrac{ m_{\nu_{3} } \left( m_{\nu_{2}} + m_{\nu_{1}} 
    - m_{\nu_{3}} + R_{\nu} \right) }{ \left( m_{\nu_{2}} - m_{\nu_{1}} 
    \right) \left( m_{\nu_{2}} - m_{\nu_{1}} - m_{\nu_{3}} + R_{\nu}\right)}} 
    & 0 \\ 
   - \sqrt{ \dfrac{ m_{\nu_{2} } - m_{\nu_{1}} + m_{\nu_{3}} - R_{\nu} 
    }{ 2\left( m_{\nu_{2}} - m_{\nu_{1}} \right) } } & 
   \sqrt{ \dfrac{ m_{\nu_{2}} - m_{\nu_{1}} - m_{3\nu} + R_{\nu} } { 
   2\left( m_{\nu_{2}} - m_{\nu_{1}} \right) } } & 0 \\ 
    0 & 0 & 1
  \end{pmatrix} .  
 \end{align}}
 Therefore, the ${\bf M}_{\nu}$ neutrino mass matrix is diagonalized by follow 
 unitary matrix 
 \begin{equation}
  {\bf U}_{\nu} = {\bf \textit{u} }_{\pi/4} {\bf P}^{\dagger}_{\nu} 
  {\bf O}_{\nu}.
 \end{equation}
 Then, one obtains that leptonic mixing matrix takes the form
 \begin{equation}
  {\bf  V}_{PMNS} = {\bf U}^{\dagger}_{\ell} {\bf U}_{\nu} 
  = {\bf O}^{T}_{\ell L} {\bf Q}^{\dagger}_{\ell} {\bf S}_{23} 
  {\bf P}^{\dagger}_{\nu} {\bf O}_{\nu}.
 \end{equation}
 Here, 
 ${\bf S}_{23} = {\bf U}^{T}_{ \pi/4} {\bf \textit{u} }_{\theta = \pi/4}$ is 
 the permutation matrix which is an element of the $S_{3}$ family group, at 
 the same time, it is an element of the $Q_{6}$ family since $S_{3}$ is a 
 subgroup of it. Therefore, the PMNS mixing matrix has the following form
{\small
 \begin{align}\label{eq23}
  {\bf V}_{PMNS}= 
  \begin{pmatrix}
   O_{11 \ell} O_{11 \nu} + O_{31\ell} O_{21\nu} e^{ i \bar{\eta}_{3 e} }  & 
   O_{11 \ell} O_{12 \nu} + O_{31\ell} O_{22\nu} e^{ i \bar{\eta}_{3 e} }  & 
   O_{21 \ell} e^{ i\eta_{2 e} } \\ 
   O_{12 \ell} O_{11 \nu} + O_{32 \ell} O_{21 \nu} e^{ i \bar{\eta}_{3 e} } & 
   O_{12 \ell} O_{12 \nu} + O_{32\ell} O_{22\nu}   e^{ i \bar{\eta}_{3 e} } & 
   O_{22 \ell} e^{i \eta_{2 e} }  \\ 
   O_{13 \ell} O_{11 \nu} + O_{33 \ell} O_{21 \nu} 
    e^{ i \bar{\eta}_{3 e} }  &  
   O_{13 \ell} O_{12 \nu} + O_{33\ell} O_{22\nu} e^{ i \bar{\eta}_{3 e} } & 
   O_{23\ell} e^{ i \eta_{2 e} } ,
  \end{pmatrix} 
 \end{align}}
 where $\eta_{2 e}$ and $\bar{\eta}_{3 e}$ are phases that coming from lepton 
 mass matrices. 
 In this case the $V_{PMNS}$ mixing matrix has three free parameters which are 
 $\eta_{2 e}$, $\bar{\eta}_{3 e}$, $y_{l}$. Since the charged lepton mass ratios are given by
 experimental data~\cite{Nakamura:2010zzi}, while the neutrino masses are determined by means of the  
 sum rule, Eq.~(\ref{eqsum}), and the neutrino mass squared splittings, Eq~(\ref{data}). 
  Comparing this matrix with the {\it standard parametrization} 
 given in~\cite{Beringer:1900zz}, we find that the reactor, the atmospheric  
 and the solar mixing angles are well determined as follows
{\small
 \begin{align}
  \vert \sin \theta_{13}^{\ell^{th}} \vert = \vert O_{21\ell} \vert , \quad 
  \vert \sin \theta_{23}^{\ell^{th}} \vert = \dfrac{ \vert O_{22\ell} \vert 
   }{ \sqrt{ 1 - \vert O_{21\ell} \vert^2} } ,\quad
  \vert \tan \theta_{12}^{\ell^{th}} \vert^2 = 
   \dfrac{ \vert O_{11 \ell} O_{12 \nu} + O_{31 \ell} O_{22\nu} 
   e^{ i \bar{\eta}_{3 e} } \vert^2  }{ \vert O_{11 \ell} 
   O_{11 \nu} + O_{31 \ell } O_{21 \nu} 
   e^{ i \bar{\eta}_{3 e} } \vert^2 }.\label{eq24}
 \end{align}}
 Let us point out a remarkable coincidence between the above formulas and 
 those showed in~\cite{Felix:2006pn,Mondragon:2007af}, their functional 
 behaviour seems to be the same, at least. As we already commented briefly, it 
 is not a surprise since the $Q_{6}$ family group is the double covering 
 of the $S_{3}$ one, so that  in this particular model the ${\bf S}_{23}$ 
 presence in the leptonic sector is not simply a coincidence. Of course, we 
 expect that our results turn out being different to the $S_3$ case, since the 
 charged lepton and neutrino contributions are different in both models, as we 
 will see next. 

\subsubsection{The masses of the right-handed neutrinos without degeneration}  
 Now, we consider the case where the RHN masses are not degenerate. 
 In this case we can have both hierarchies of the neutrino masses, 
 as we will. The effective neutrino mass matrix 
 ${\bf m}_{\nu}$ can be written in polar form as  
 $P_{\nu} \bar{{\bf m}}_{\nu} P_{\nu}$, where $\bar{\bf m}_{\nu}$ is a 
 symmetric real matrix and 
 $P_{\nu} = \textrm{diag}\left( 1, e^{i \alpha_{1} }, e^{i \alpha_{2} } 
 \right)$ is a diagonal matrix  of phases with 
 $ 2\alpha_{1} = \arg\{\mu_{0}\} - \arg\{b_{\nu}\} $, 
 $ 2\alpha_{2} = \arg\{ d_{\nu} \} - \arg\{b_{\nu}\} $, 
 $ 2\arg\{c_{\nu}\} = \arg\{b_{\nu}\} + \arg\{d_{\nu}\}$ and 
 $ 2\arg\{a_{\nu}\} = \arg\{b_{\nu}\} + 
 \arg\{\mu_{0}\}$~\cite{Canales:2011ug,Canales:2012dr}. The 
 symmetric real matrix with one texture zero, $\bar{\bf m}_{\nu}$, can be 
 expressed in terms of a matrix with two texture zeros class~I 
 as~\cite{Canales:2012dr}: 
 \begin{equation}
  \bar{\bf m}_{\nu} = \mu_{0} \mathbb{I}_{3 \times 3} + {\bf M}'_{\nu}
 \end{equation}
 where the matrix ${\bf M}'_{\nu}$ written in terms of its eigenvalues, for a 
 normal [inverted] hierarchy, is~\cite{Canales:2013cga}:
 \begin{equation}
  \widetilde{\bf M}'_{\nu} = \frac{ {\bf M}'_{\nu} }{ \sigma_{3[2]} } 
  \left( \begin{array}{ccc} \vspace{2mm}
   \widetilde{\sigma}_{1[3]} - \widetilde{\sigma}_{2[1]} + \delta_{\nu} & 
   \sqrt{ \frac{ \widetilde{\sigma}_{1[3]} \widetilde{\sigma}_{2[1]} }{ 1 - 
   \delta_{\nu} } } &
   \sqrt{ \frac{ \delta_{\nu} }{ 1 - \delta_{\nu} } f_{\nu 1[3] } f_{\nu 2[1] 
    }  } \\ \vspace{2mm}
   \sqrt{ \frac{ \widetilde{\sigma}_{1[3]} \widetilde{\sigma}_{2[1]} }{ 1 - 
   \delta_{\nu} } } &  
   0 & 0 \\ \vspace{2mm}
   \sqrt{ \frac{ \delta_{\nu} }{ 1 - \delta_{\nu} } f_{\nu 1[3] } f_{\nu 2[1]}  
   } & 0 & 1 - \delta_{\nu}
  \end{array} \right),
 \end{equation}
 where
 \begin{equation}\label{Der:Fs}
  \begin{array}{l}\vspace{2mm}
   f_{\nu 1[3]} = \left( 1 - \widetilde{\sigma}_{1[3]} - \delta_{\nu} \right), 
    \qquad
   f_{\nu 2[1]} = \left( 1 + \widetilde{\sigma}_{2[1]} - \delta_{\nu} \right),  
    \\\vspace{2mm}
   \widetilde{\sigma}_{1[3]} = \frac{ \widetilde{m}_{\nu_{1[3]}} - 
    \widetilde{\mu} }{ 1 - \widetilde{\mu}}, \quad 
   \widetilde{\sigma}_{2[1]} = \frac{ \left| \widetilde{m}_{\nu_{2[1]}} - 
    \widetilde{\mu} \right| }{ 1 - \widetilde{\mu} } , \quad
   \widetilde{m}_{\nu_{1[3]}} = \frac{ m_{\nu_{1[3]}} }{ m_{\nu_{3[2]}} }, 
    \quad
   \widetilde{m}_{\nu_{2[1]}} = \frac{ m_{\nu_{2[1]}} }{ m_{\nu_{3[2]}} }, 
    \quad
   \widetilde{\mu} = \frac{ |\mu_{0}| }{ m_{\nu_{3[2]}} }. 
  \end{array}
 \end{equation}
 From the expressions for the mass parameters, eq.~(\ref{Der:Fs}), we
 obtain the following constraint 
 $\widetilde{m}_{\nu_{1[3]}} > \widetilde{\mu} $. The parameter
 $\delta_{\nu}$ is defined as $\delta_{\nu} = 1 - \widetilde{d}_{\nu}
 + \widetilde{\mu}$ with $\widetilde{d}_{\nu} = | d_{\nu}
 |/\sigma_{3[2]}$, and has a  range of values $1 -
 \widetilde{\sigma}_{\nu_{1[3]}} > \delta_{\nu} > 0$, which is
 equivalent to $m_{\nu_{3[2]}} > |d_{\nu}| > m_{\nu_{1[3]}}$. From the
 above it can be seen that strictly speaking $|d_{\nu}|$ is not a
 free parameter of the  mass matrix ${\bf M}_{\nu}$, and consequently of
 ${\bf m}_{\nu}$, because it must meet the above condition. The
 numerical values consistent with this condition are determined from the
 experimental data on neutrino oscillations. Now, reparameterized in terms of its
 eigenvalues, the orthogonal real matrix that diagonalizes the mass
 matrix $\widetilde{\bf M}'_{\nu}$ (and thus $\bar{\bf
   m}_{\nu}$) is:
 \begin{equation}\label{Onu:noDeg}
  {\bf O}_{\nu} = 
  \left( \begin{array}{ccc} \vspace{2mm}
  \sqrt{ \frac{ \widetilde{\sigma}_{1[3]} \left( 1 - \delta_{\nu} \right) 
   f_{\nu 1[3]} }{ D_{\nu 1[3]} } } &
  \sqrt{ \frac{ \widetilde{\sigma}_{2[1]} \left( 1 - \delta_{\nu} \right) 
   f_{\nu 2[1]} }{ D_{\nu 2[1]} } } &
  \sqrt{ \frac{ \delta_{\nu} \left( 1 - \delta_{\nu} \right) }{ D_{ \nu 3[2]} 
    } } \\ \vspace{2mm}
  \sqrt{ \frac{ \widetilde{\sigma}_{2[1]} f_{\nu 1[3]} }{ D_{\nu 1[3] } } } & 
  -\sqrt{ \frac{ \widetilde{\sigma}_{1[3] } f_{\nu 2[1] } }{ D_{\nu 2[1] }} } 
   &
  \sqrt{ \frac{ \widetilde{\sigma}_{1[3] } \widetilde{\sigma}_{2[1] } 
   \delta_{\nu} }{ D_{\nu 3[2] } } } \\ \vspace{2mm}
  -\sqrt{ \frac{ \widetilde{\sigma}_{1[3]} \delta_{\nu} f_{\nu 2[1] } }{ 
   D_{\nu 1[3] } } } & 
  -\sqrt{ \frac{ \widetilde{\sigma}_{2[1]} \delta_{\nu} f_{\nu 1[3] } }{ 
   D_{\nu 2[1] } } } &
   \sqrt{ \frac{ f_{\nu 1[3] } f_{\nu 2[1] } }{ D_{\nu 3[2] } } } 
  \end {array}  \right),
 \end{equation}
 where
 \begin{equation}
  \begin{array}{l}\vspace{2mm}
   D_{\nu 1[3]} = \left( 1 - \delta_{\nu} \right) 
    \left( \widetilde{\sigma}_{1[3]} + \widetilde{\sigma}_{2[1]} \right) 
    \left( 1 - \widetilde{\sigma}_{1[3]} \right), \\ \vspace{2mm}
   D_{\nu 2[1]} = \left( 1 - \delta_{\nu} \right) 
    \left( \widetilde{\sigma}_{1[3]} + \widetilde{\sigma}_{2[1]} \right) 
    \left( 1 + \widetilde{\sigma}_{2[1]} \right), \\ \vspace{2mm}
   D_{\nu 3[2]} = \left( 1 - \delta_{\nu} \right) 
    \left( 1 - \widetilde{\sigma}_{1[3]} \right) 
    \left( 1 + \widetilde{\sigma}_{2[1]} \right).
  \end{array}
 \end{equation}
 Then, one obtains that the leptonic mixing matrix takes the form
 \begin{equation}
  {\bf  V}_{PMNS} = {\bf  O}^{T}_{\ell L} {\bf Q}^{\dagger}_{\ell} 
   {\bf S}_{23} {\bf P}^{\dagger}_{\nu} {\bf O}_{\nu } .
 \end{equation}
 In here ${\bf S}_{23} = {\bf U}^{T}_{ \pi/4} {\bf \textit{u} }_{\theta }$ 
 is not the permutation matrix show in the previous section, its  explicit form is 
 \begin{equation}
  {\bf S}_{23} = 
  \left(\begin{array}{ccc}
   S_{\theta + \frac{\pi}{4} } & 0 & -C_{\theta + \frac{\pi}{4} } \\
   C_{\theta + \frac{\pi}{4} } & 0 & S_{\theta + \frac{\pi}{4} } \\
   0 & 1 & 0 
  \end{array}    \right),
 \end{equation}  
 where 
 $S_{\theta + \frac{\pi}{4}} = \sin\left( \theta + \frac{\pi}{4} \right)$ and 
 $C_{\theta + \frac{\pi}{4}} = -\cos\left( \theta + \frac{\pi}{4} \right)$, 
with {\small$\theta = \arctan \left\{ \frac{ M_{R_{2}} }{ M_{R_{1}} } \right\} $}.
 Thus, the parameter $\theta$ measures the degeneracy between the first two right-handed neutrino 
 masses, and may give us a hint about the hierarchy that obeys the mass spectrum of right-handed 
 neutrinos.Therefore, the PMNS mixing matrix has the following form
 \begin{equation}\label{eq23'}
  {\bf V}_{PMNS} = 
   \begin{pmatrix}
   V_{e1}     & V_{e2}     & V_{e3} \\
   V_{\mu 1}  & V_{\mu 2}  & V_{\mu 3} \\
   V_{\tau 1} & V_{\tau 2} & V_{\tau 3}
  \end{pmatrix},
 \end{equation} 
 where 
 {\footnotesize 
 \begin{equation}
  \begin{array}{l}\vspace{2mm}
   V_{e1} =  
    \left( O_{11\ell} S_{\frac{\pi}{4} +\theta} + O_{21\ell} C_{\frac{\pi}{4} 
    + \theta} e^{i\eta_{1}} \right) O_{11\nu} + O_{31 \ell} O_{21 \nu} 
    e^{i\bar{ \eta}_{3e} } + \left( O_{21\ell} S_{\frac{\pi}{4} +\theta} 
    e^{i\eta_{e2}} - O_{11\ell} C_{\frac{\pi}{4} +\theta} 
    e^{i\alpha_{2}} \right) O_{31\nu}  \\ \vspace{2mm}
   V_{e2} =  \left( O_{11\ell} S_{\frac{\pi}{4} +\theta} + O_{21\ell} 
    C_{\frac{\pi}{4} +\theta} e^{i\eta_{1}} \right) O_{12\nu} + O_{31 \ell} 
    O_{22 \nu} e^{i\bar{ \eta}_{3e} } + \left( O_{21\ell} S_{\frac{\pi}{4} 
    +\theta} e^{i\eta_{e2}} - O_{11\ell} C_{\frac{\pi}{4} +\theta} 
    e^{i\alpha_{2}} \right) O_{32\nu}  \\ \vspace{2mm} 
   V_{e3} =  \left( O_{11\ell} S_{\frac{\pi}{4} +\theta} + O_{21\ell} 
    C_{\frac{\pi}{4} +\theta} e^{i\eta_{1}} \right) O_{13\nu} + O_{31 \ell} 
    O_{23 \nu} e^{i\bar{ \eta}_{3e} } + \left( O_{21\ell} S_{\frac{\pi}{4} 
    +\theta} e^{i\eta_{e2}} - O_{11\ell} C_{\frac{\pi}{4} +\theta} 
    e^{i\alpha_{2}} \right) O_{33\nu}  \\ \vspace{2mm}
   V_{\mu 1} = \left( O_{12\ell} S_{\frac{\pi}{4} +\theta} + O_{22\ell} 
    C_{\frac{\pi}{4} +\theta} e^{i\eta_{1}} \right) O_{11\nu} + O_{32 \ell} 
    O_{21 \nu} e^{i\bar{ \eta}_{3e} } + \left( O_{22\ell} S_{\frac{\pi}{4} 
    +\theta} e^{i\eta_{e2}} - O_{12\ell} C_{\frac{\pi}{4} +\theta} 
    e^{i\alpha_{2}} \right) O_{31\nu}  \\ \vspace{2mm} 
   V_{\mu 2} = \left( O_{12\ell} S_{\frac{\pi}{4} +\theta} + O_{22\ell} 
    C_{\frac{\pi}{4} +\theta} e^{i\eta_{1}} \right) O_{12\nu} + O_{32 \ell} 
    O_{22 \nu} e^{i\bar{ \eta}_{3e} }
    + \left( O_{22\ell} S_{\frac{\pi}{4} +\theta} e^{i\eta_{e2}} - O_{12\ell} 
     C_{\frac{\pi}{4} +\theta} 
     e^{i\alpha_{2}} \right) O_{32\nu}  \\ \vspace{2mm}  
   V_{\mu 3} = \left( O_{12\ell} S_{\frac{\pi}{4} +\theta} + O_{22\ell} 
    C_{\frac{\pi}{4} +\theta} e^{i\eta_{1}} \right) O_{13\nu} + O_{32 \ell} 
    O_{23 \nu} e^{i\bar{ \eta}_{3e} }
    + \left( O_{22\ell} S_{\frac{\pi}{4} +\theta} e^{i\eta_{e2}} - O_{12\ell} 
    C_{\frac{\pi}{4} +\theta} 
    e^{i\alpha_{2}} \right) O_{33\nu}  \\ \vspace{2mm} 
   V_{\tau 1} = \left( O_{13\ell} S_{\frac{\pi}{4} +\theta} + O_{23\ell} 
    C_{\frac{\pi}{4} +\theta} e^{i\eta_{1}} \right) O_{11\nu} + O_{33 \ell} 
    O_{21 \nu} e^{i\bar{ \eta}_{3e} }
    + \left( O_{23\ell} S_{\frac{\pi}{4} +\theta} e^{i\eta_{e2}} - O_{13\ell} 
    C_{\frac{\pi}{4} +\theta} 
    e^{i\alpha_{2}} \right) O_{31\nu}  \\ \vspace{2mm}    
   V_{\tau 2} = \left( O_{13\ell} S_{\frac{\pi}{4} +\theta} + O_{23\ell} 
    C_{\frac{\pi}{4} +\theta} e^{i\eta_{1}} \right) O_{12\nu} + O_{33 \ell} 
    O_{22 \nu} e^{i\bar{ \eta}_{3e} }
    + \left( O_{23\ell} S_{\frac{\pi}{4} +\theta} e^{i\eta_{e2}} - O_{13\ell} 
    C_{\frac{\pi}{4} +\theta} 
    e^{i\alpha_{2}} \right) O_{32\nu}  \\ \vspace{2mm}
   V_{\tau 3} = \left( O_{13\ell} S_{\frac{\pi}{4} +\theta} + O_{23\ell} 
    C_{\frac{\pi}{4} +\theta} e^{i\eta_{1}} \right) O_{13\nu} + O_{33 \ell} 
    O_{23 \nu} e^{i\bar{ \eta}_{3e} }
    + \left( O_{23\ell} S_{\frac{\pi}{4} +\theta} e^{i\eta_{e2}} - O_{13\ell} 
    C_{\frac{\pi}{4} +\theta} e^{i\alpha_{2}} \right) O_{33\nu}       
  \end{array}
 \end{equation}  }
 with $\eta_{2e}= \eta_{1} + \alpha_{2}$ and 
 $\bar{\eta}_{3e} = \eta_{2} + \alpha_{1}$. 
Now, for this case the $V_{PMNS}$ mixing matrix 
 has nine free parameters which are $y_{\ell}$, $\theta$, $\eta_{1}$, $\alpha_{2}$, $\eta_{2e}$, 
 $\bar{\eta}_{3e}$, $m_{\nu_{3[2]}}$, $\widetilde{\mu}_{0}$ and $\delta_{\nu}$.
 From the above expressions it is easy to see that the case analysed in the previous 
 section is a particular case of the one discussed in this section. In other 
 words, if $\theta = \pi/4$ we have $\tan \theta = 1$ 
 which implies $M_{R_{1}} = M_{R_{2}}$.

\section{Numerical analysis for the mixing matrices} 
 
\subsection{CKM mixing matrix}
 The CKM matrix is defined as ${\bf V}_{CKM}=
 {\bf U}^{\dagger}_{u L} {\bf U}_{d L} = {\bf O}^{T}_{u L} {\bf Q}_{q}
 {\bf O}_{d L}$, with the form of the mass matrices found in the
 previous sections. So far, there are three free parameters ($y_{d}$ and
 two CP-violating phases in ${\bf Q}_{q}$) if the quark mass ratios are
 taken as inputs. As it is well known, the physical masses depend on the
 scale at which they are measured,  in this model the CKM matrix 
 may be obtained numerically with masses at the GUT scale. However, the mass 
 ratios do not change drastically at different energy scales as one can verify 
 directly from~\cite{Xing:2007fb}. Therefore, we will assume that the form of 
 the mass matrices will remain the same from the GUT scale to the electroweak 
 scale. Of course, in the more detailed analysis that is currently in 
 progress, the effects of the extra Higgs fields in the model and the running 
 of the renormalization group will be taken into account. 
 Thus, for the rest of the analysis we will assume we are already at the
 electroweak scale.  At low energies,  we have used the following values for the quark mass ratios  
 given in~\cite{Canales:2012ix, Canales:2013cga}:
 \begin{equation}
  \begin{array}{l}
   m_{u}/m_{t} = (1.73 \pm 0.75) \times 10^{-5}, \quad 
   m_{c}/m_{t} = (3.46 \pm 0.43) \times 10^{-3}, \\
   m_{d}/m_{b} = (1.12 \pm 0.007) \times 10^{-3}, \quad 
   m_{s}/m_{b}= (2.32 \pm 0.84) \times 10^{-2}. 
  \end{array}
 \end{equation}
 
 In order to obtain the numerical values for the three free parameters,
 we perform a $\chi^{2}$ analysis on the parameter space, to find their
 best fit points. It is built as in~\cite{Canales:2012ix, Canales:2013cga}

 \begin{equation}
  \chi^2= \dfrac{ ( |V^{th}_{ud}| - |V^{ex}_{ud}|)^2 } { \sigma^{2}_{V_{ud}} } 
   + \dfrac{ ( |V^{th}_{us}| - |V^{ex}_{us}| )^2 }{ \sigma^{2}_{V_{us} } } 
   + \dfrac{ ( |V^{th}_{ub}| - |V^{ex}_{ub}| )^2 }{ \sigma^{2}_{V_{ub}} } 
   + \dfrac{ ( \mathcal{J}^{th}_{q} - \mathcal{J}^{ex}_{q})^2 }{ 
   \sigma^{2}_{\mathcal{J}_{q}}}
 \end{equation}
 where we have taken the following experimental values for the
 $V_{CKM}$ elements we used to construct the $\chi^2$ function~\cite{Beringer:1900zz}:
 \begin{equation}
  \begin{array}{l}
   \left| V^{ex}_{ud} \right| = 0.97427 \pm 0.00015, \quad
   \left| V^{ex}_{us} \right| = 0.2253 \pm 0.007, \\
   \left| V^{ex}_{ub} \right| = 0.00351 \pm 0.00015, \quad 
   \mathcal{J}^{ex}_{q} = (2.96 \pm 0.18) \times 10^{-5}.
  \end{array}
 \end{equation}
 Notice that using the Jarlskog invariant in the $\chi^2$ function implies 
 unitarity as a constraint. The best values for the free parameters are thus 
 found to be
 \bea
  y_{d}  = 0.981977^{+0.002843 }_{-0.002117}, \quad
  \alpha = \left( 168.78^{+2.10}_{-1.11} \right)^{\circ}, \quad
  \beta  = \left( 132^{+105}_{-25} \right)^{\circ} ,
 \eea
 at 70 \% C. L with $\chi^{2} = 0.0515$ as the minimal value.
 These correspond to the following values for the $V_{CKM}$ elements
 \begin{eqnarray}
  |V^{th}_{ud}| & = & 0.97428^{+0.00016}_{-0.00014},\quad 
  |V^{th}_{us}| = 0.2252^{+0.00062}_{-0.00067} \nonumber \\
  |V^{th}_{ub}| & = & 0.00351^{+0.00017}_{-0.00016}, \quad 
  \mathcal{J}_{q} = 2.95^{+0.20}_{-0.19} \times 10^{-5}. \label{eq.obser}
 \end{eqnarray}

 \begin{figure}[ht]
  \centering
  \includegraphics[scale=0.45]{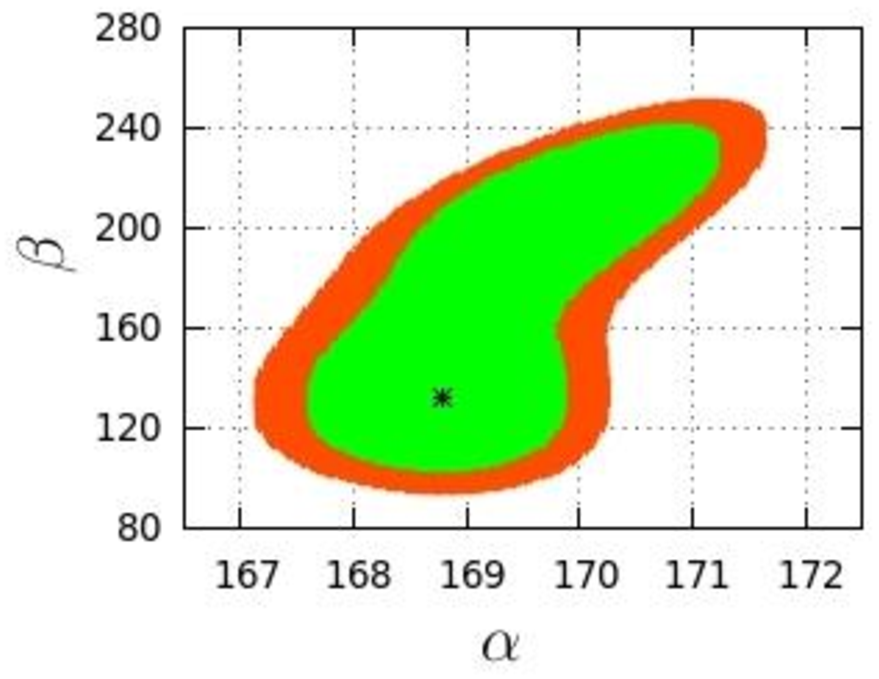}\hspace{-0.5cm}\includegraphics[scale=0.45]{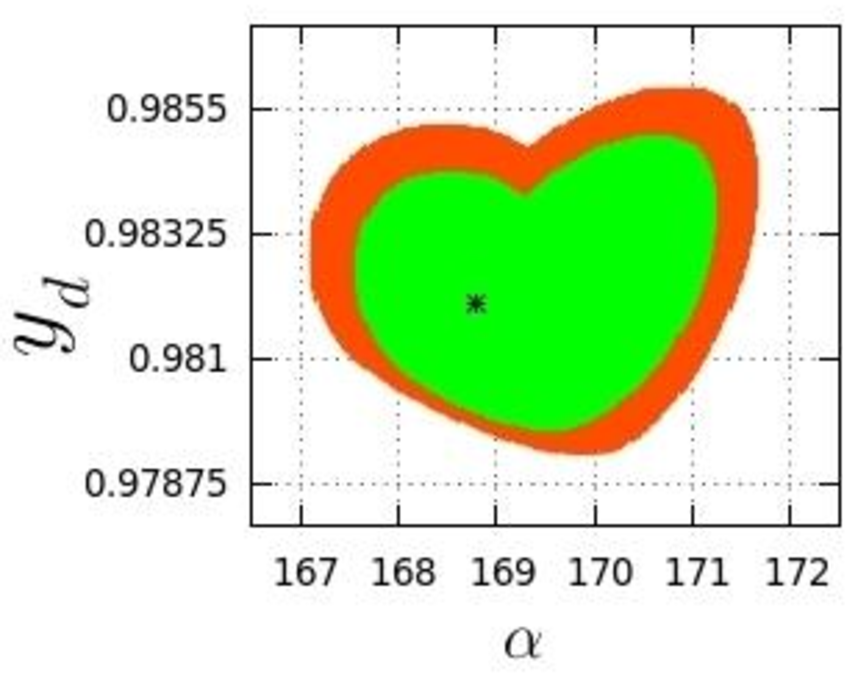}\includegraphics[scale=0.45]{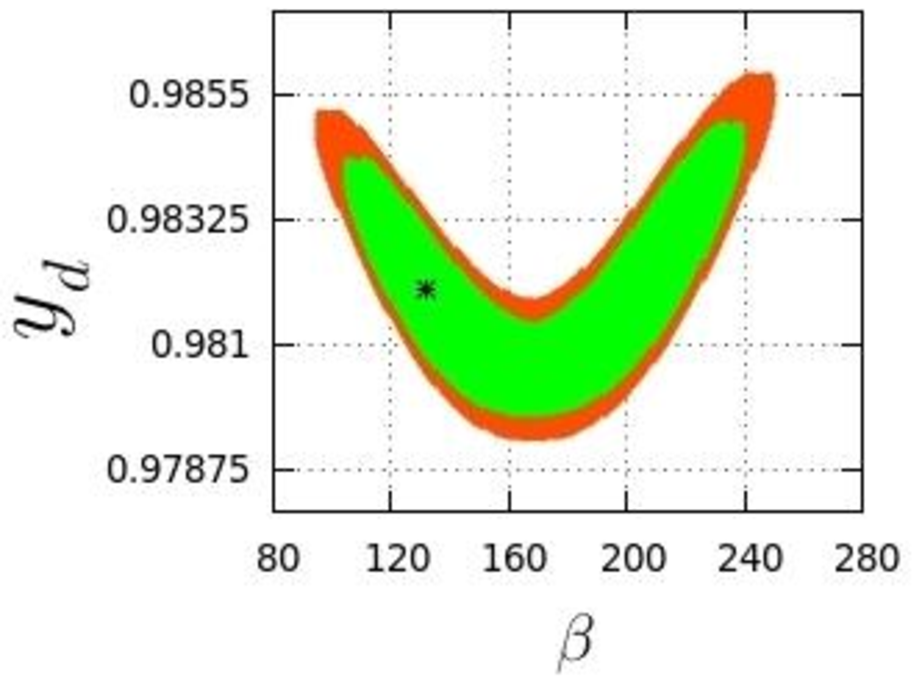}
   \caption{Allowed region for the three free parameters in 
    the quark sector at $70\%$ (green) and $90\%$ 
   (orange)  confidence level.}
  \label{eq29.1}
 \end{figure}

\subsection{PMNS mixing matrix}

\subsubsection{The masses of right-handed neutrinos with degeneration}
 As we observe from  Eq.~(\ref{eq24}), the reactor and atmospheric
 angles turn out to be independent of the neutrino masses. 
 These observables only depend explicitly on the $y_{\ell}$ free parameter, the 
 charged lepton masses and the $\eta_{2 e}$ Dirac phase; the latter may be 
 ignored since we are only interested in the absolute values of the two mixing 
 angles. On the other hand, the solar mixing angle depends on the $y_{\ell}$ free 
 parameter, the $\bar{\eta}_{3e}$ phase, as well as the charged lepton and 
 neutrino masses. 
 
 In order to show that in this model we can describe the lepton masses and
 mixing  we will also make a $\chi^{2}$ fit using the 
 theoretical expressions for the atmospheric and reactor angles given in 
 Eq.~(\ref{eq24}) and compare them with the current experimental data for 
 these mixing angles. Thus, for the atmospheric mixing angle we consider the 
 following experimental value 
 $\sin^2\theta^{\ell^{ex}}_{23}= 0.52 \pm 0.06$. Within this 
 theoretical framework we will consider first the particular case when the first and 
 second  right-handed neutrinos are mass degenerate. In this case we can only determine 
 a lower bound for the value of the reactor angle~\cite{Canales:2012dr}. 
 To perform the  $\chi^{2}$ fit we considered the values for the 
 reactor angle reported by the MINOS experiment; 
 $\sin^2 2\theta^{\ell^{ex}}_{13}=0.076\pm 0.068$. We considered also the following  
 charged lepton masses values: $m_{e}=0.51099$~MeV, $m_{\mu}= 105.6583$~MeV 
 and $m_{\tau}=1776.82$~MeV~\cite{Nakamura:2010zzi}. As a result of this
 $\chi^{2}$ analysis, we obtained that for the best fit with $\chi^2=0.85$ as 
 the minimum value, the free parameter $y_{\ell}$ at $1\sigma$ has the following 
 range of values: $y_{\ell}= 0.8478^{+0,0045}_{-0.0046}$. Moreover, the best 
 theoretical values for the atmospheric and reactor angles that come from this 
 analysis, at $1\sigma$, are: 
 \begin{equation}\label{eqang}
  \begin{array}{l}\vspace{2mm}
   \sin^{2} \theta^{\ell^{th}}_{23} = 0.5206_{-0.0113}^{+0.0115} 
    \rightarrow \theta^{\ell^{th}}_{23} = \left( 46.18^{+0.66}_{-0.65} \right)^{\circ} , \\ 
   \sin^{2} 2 \theta^{\ell^{th}}_{13} = 0.01386_{-0.00025}^{+0.00016}   
    \rightarrow \theta^{\ell^{th}}_{13} = \left( 3.38^{+0.03}_{-0.02} \right)^{\circ} .
 \end{array}
 \end{equation}
  
\begin{figure}[ht]
  \centering
   \includegraphics[scale=.7]{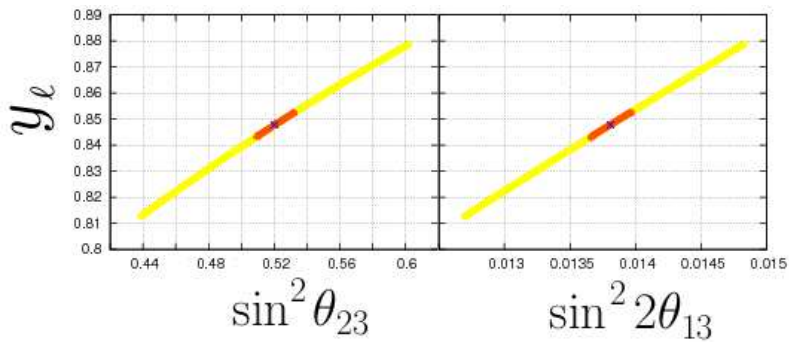}
   \caption{Atmospheric and reactor mixing angle values at $65\%$ (orange) 
   and $95\%$  (yellow) of C.L.}
  \label{eq29.2}
 \end{figure}

 As can be observed, the atmospheric angle value is in good agreement
 with the experimental values, however, the obtained reactor mixing
 value is smaller than the central value obtained in the global fits. In fact, this theoretical
 value of the reactor angle is more than $3 \sigma$ away from the
 values reported in ref.~\cite{Tortola:2012te,GonzalezGarcia:2012sz}.
 But it is worth noting that our model predicts that the reactor angle
 must be different from zero and that is large in comparison to the
 one obtained from the tribimaximal mixing matrix. Moreover, as
 already stated above, it is a lower bound for the value obtained in a
 more general model where the right-handed neutrino masses are not
 degenerate. In figure \ref{eq29.2}, we show the dependence of the
 atmospheric and reactor angles on the parameter $y_{\ell}$.
 
\begin{figure}[ht]
  \centering
    \includegraphics[scale=0.5]{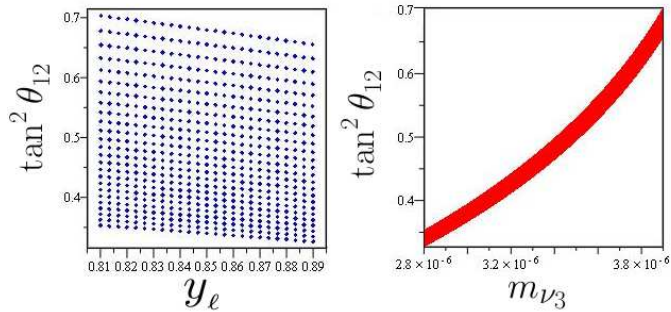}
   \caption{Solar mixing angle values as function of the $y_{\ell}$ parameter and 
    the $m_{\nu_{3}}$ neutrino mass considering $m_{\nu_{2}}= 0.05080$~eV,  
    $m_{\nu_{1}}=0.04987$~eV and $\bar{\eta}_{3 e}=\pi$. } \label{eq29.3}
 \end{figure}

 Having determined the allowed values for the $y_{\ell}$ free parameter, then, 
 the solar mixing angle just depends on the neutrino masses and one Dirac 
 phase. We can reduce further the free parameters, noticing that the neutrino 
 masses can be determined using the sum rule given in Eq.~(\ref{eqsum}). 
 This is written in terms of the observables $\Delta m^{2}_{\odot}$ and 
 $\Delta m^{2}_{ATM}$ as
 \beq \label{newsum}
   m_{\nu_{3}} \leq \left( \sqrt[4]{ m^{2}_{\nu_{3}} + \Delta m^{2}_{\odot} 
   + \Delta m^{2}_{ \textrm{ATM} } } 
  - \sqrt[4]{ m^{2}_{\nu_{3}} + \Delta m^{2}_{ \textrm{ATM} }} \right)^2.
 \eeq
 From the above expression and using the experimental results of 
 $\Delta m^{2}_{\odot}$ and $\Delta m^{2}_{ATM }$, we can get an upper bound 
 for the $m_{\nu_{3}}$ lightest neutrino mass. Therefore, the allowed values 
 for the lightest neutrino mass are: 
 $0\leq m_{\nu_{3}}\leq 4\times 10^{-6}$~eV. As a result, the $m_{\nu_{2}}$ 
 and $m_{\nu_{1}}$ neutrino masses are easily calculated in the following way
 \bea \label{masses} m_{\nu_{2}} & = & \sqrt{ \Delta m^{2}_{\odot} +
   \Delta m^{2}_{ \textrm{ATM} } } \sqrt{ 1 + \dfrac{m^{2}_{\nu_{3}}
   }{ \Delta m^{2}_{\odot} + \Delta m^{2}_{ \textrm{ATM} } } }
 \approx 5.0 \pm 0.087 \times 10^{-2}~eV,\nn\\
 m_{\nu_{1}} & = & \sqrt{ \Delta m^{2}_{ \textrm{ATM} } } \sqrt{ 1 +
   \dfrac{m^{2}_{\nu_{3}} }{ \Delta m^{2}_{ \textrm{ATM} } } } \approx
 4.90 \pm 0.089 \times 10^{-2}~eV.  \eea 
 As we already commented, this version of our model where the first two
 RHN masses are degenerate, predicts an inverted ordering among
 the neutrino masses, the above values clarify explicitly our
 statement. Since we have calculated the neutrino masses, these cease
 to be considered as free parameters in the solar mixing angle
 expression given in Eq.~(\ref{eq24}).  Furthermore, we obtain easily
 the solar mixing value that is allowed by the fixed free parameters,
 which are $y_{\ell}$ and the neutrino masses. Actually, with the
 particular value for the $\bar{\eta}_{3 e}=\pi$ Dirac phase, and
 using the $y_{\ell}$ value at $90\%$ at C.L, we predict the following value 
 for the solar mixing angle 
 \beq \label{eqsolar} 
  \tan^{2}  \theta^{\ell^{th}}_{12} = 0.552 \pm 0.078 \rightarrow
  \theta^{\ell^{th}}_{12} = \left( 36.62 \pm 4.06  \right)^{\circ}
 \eeq 
 where we have used $m_{\nu_{2}}= 0.05080$~eV,
 $m_{\nu_{1}}=0.04987$~eV and $m_{\nu_{3}}=3.9\times 10^{-6}$~eV. As
 we observe the solar mixing angle value in Eq.~(\ref{eqsolar}) is in
 good agreement with the experimental data.  In figure \ref{eq29.3} we
 show explicitly the dependence of the solar angle on the $y_{\ell}$
 parameter and the $m_{\nu_{3}}$ lightest neutrino mass, respectively.

 It is clear from these results that the solar angle has a strong dependence 
 on the $m_{\nu_{3}}$ mass, and that the sum rule for the neutrino masses does 
 play an important role to determine the above mixing angle in good agreement 
 with the experimental data.
 
\subsubsection{The masses of right-handed neutrinos without degeneration}

 In the previous section we obtained that the reactor mixing angle, $\theta^{\ell}_{13}$, has a 
 nonzero value, whereby the shape of lepton mixing matrix PMNS is not consistent with 
 tribimaximal scenario. However, this value of the reactor mixing angle is small compared to the 
 central value from the global fits~\cite{Forero:2014bxa}, and would correspond to a lower bound. 
  
 Now, in order to reproduce the current numerical values of the reactor angle reported by  global 
 fits, we break the degeneration of the first two right-handed neutrinos, namely 
 $M_{R_{1}} \neq M_{R_{2}}$. Also, we defined the $\chi^{2}$ function as:
 \begin{equation}
  \chi^{2} = 
    \frac{ \left( S_{12}^{2,ex} - S_{12}^{2,th} \right)^{2} }{\sigma_{ S_{12}^{2,ex} }^{2} } 
   +\frac{ \left( S_{23}^{2,ex} - S_{23}^{2,th} \right)^{2} }{\sigma_{ S_{23}^{2,ex} }^{2} } 
   +\frac{ \left( S_{13}^{2,ex} - S_{13}^{2,th} \right)^{2} }{\sigma_{ S_{13}^{2,ex} }^{2} } 
  \end{equation}  
  where $S_{ij}^{2,ex(th)} \equiv \sin^{2} \theta_{ij}^{\ell}$ with
  $i,j=1,2,3$. The terms with superindex $th$ are the theoretical
  expressions obtained from Eq.~(\ref{eq23'}), while the terms with
  superindex $ex$ are the experimental data with uncertainty~$
  \sigma_{ S_{13} }$ given in Eq.~(\ref{data}).  A consequence of
  considering the most general case with non-degenerate right- handed
  neutrino masses is an increasing of the number of free parameters in
  the PMNS matrix. In a first scan of the parameter space, we find
  that the $\eta_{1}$, $\alpha_{2}$,
  $\eta_{2e}$ and $\bar{\eta}_{3e}$ phases are not correlated among
  each other nor with the parameters $m_{\nu_{3[2]}}$,
  $\widetilde{\mu}_{0}$ and $\delta_{\nu}$. Thus we consider that 
  the $\eta_{1}$, $\alpha_{2}$, $\eta_{2e}$ and
  $\bar{\eta}_{3e}$ phases vary in the range of 0 to $2\pi$. From the
  analysis done in the previous section we consider that $y_{\ell} \in
  \left[ 0.75, 0.96 \right]$. Then, to perform the $\chi^{2}$ fit the
  neutrino masses are written as:
 \begin{equation}
  \begin{array}{lll}
   \widetilde{m}_{\nu_{1}} = \dfrac{ m_{\nu_{1}} }{ m_{\nu_{3}} } =  
    \sqrt{ 1 -  \frac{ \Delta m^{2}_{ \textrm{ATM} }  }{ m_{\nu_{3}}^{2}  } }, & 
   \widetilde{m}_{\nu_{2}} = \dfrac{ m_{\nu_{2}} }{ m_{\nu_{3}} } =  
    \sqrt{ 1 -  \frac{ \Delta m^{2}_{ 32 }  }{ m_{\nu_{3}}^{2}  } }  &
    \textrm{(NH),}   \\ 
   \widetilde{m}_{\nu_{3}} = \dfrac{ m_{\nu_{3}} }{ m_{\nu_{2}} } =  
    \sqrt{ 1 -  \frac{ \Delta m^{2}_{23}  }{ m_{\nu_{2}}  } }, & 
   \widetilde{m}_{\nu_{1}} = \dfrac{ m_{\nu_{1}} }{ m_{\nu_{2}} } =  
    \sqrt{ 1 -  \frac{ \Delta m^{2}_{ \odot }  }{ m_{\nu_{2}}^{2}  } }  &
    \textrm{(IH),}  
  \end{array}  
 \end{equation}
 where $\Delta m^{2}_{32} = \Delta m^{2}_{ \textrm{ATM} } - \Delta
 m^{2}_{ \odot }$ and $\Delta m^{2}_{23} = \Delta m^{2}_{ \textrm{ATM}
 } + \Delta m^{2}_{ \odot }$. The upper [lower] row corresponds to a
 normal [inverted] hierarchy. The above expressions only have one free
 parameter which is $m_{\nu_{3[2]}}$. This parameter must satisfy the
 condition $m_{\nu_{3[2]}} < \Delta m^{2}_{ 32[\odot] }$ and $\sum_{i}
 m_{\nu_{i}} < 0.23$~eV. The last one was reported by the Planck
 collaboration~\cite{Ade:2013zuv}. Thus in the $\chi^{2}$ fit we
 consider that $m_{\nu_{3[2]}}$ is not a free parameter, since it is
 constrained by experimental data.  Thus, with $\theta = 283^{\circ}[121^{\circ}]$ for
 a normal [inverted] hierarchy, we have that  $\widetilde{\mu}_{0}$ and $\delta_{\nu}$ 
 are the only free parameters in the fit.
 
\begin{figure}[h]
  \centering
  \includegraphics[scale=0.6]{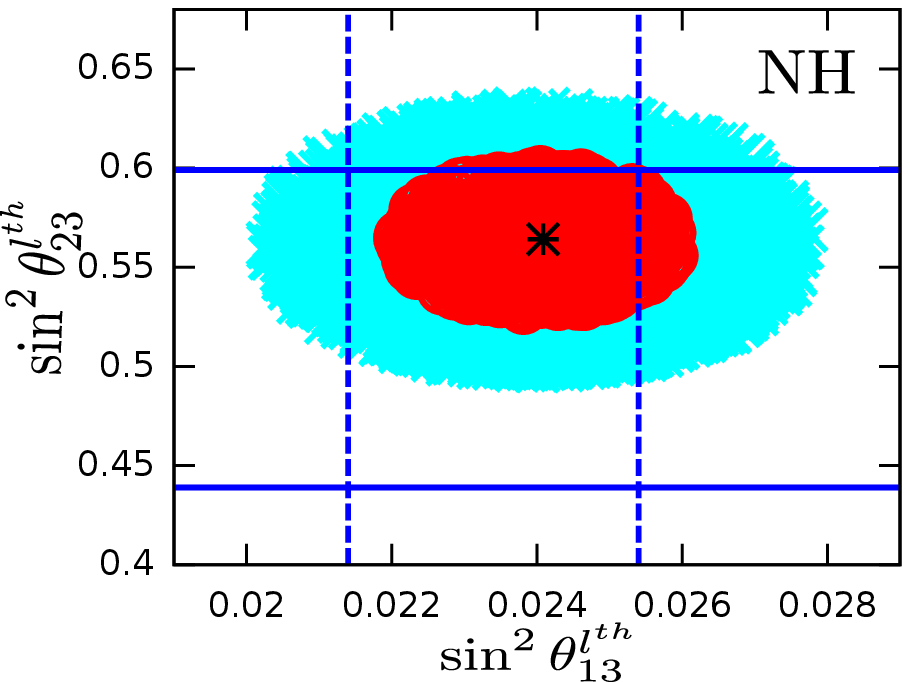}\hspace{6mm}\includegraphics[scale=0.6]{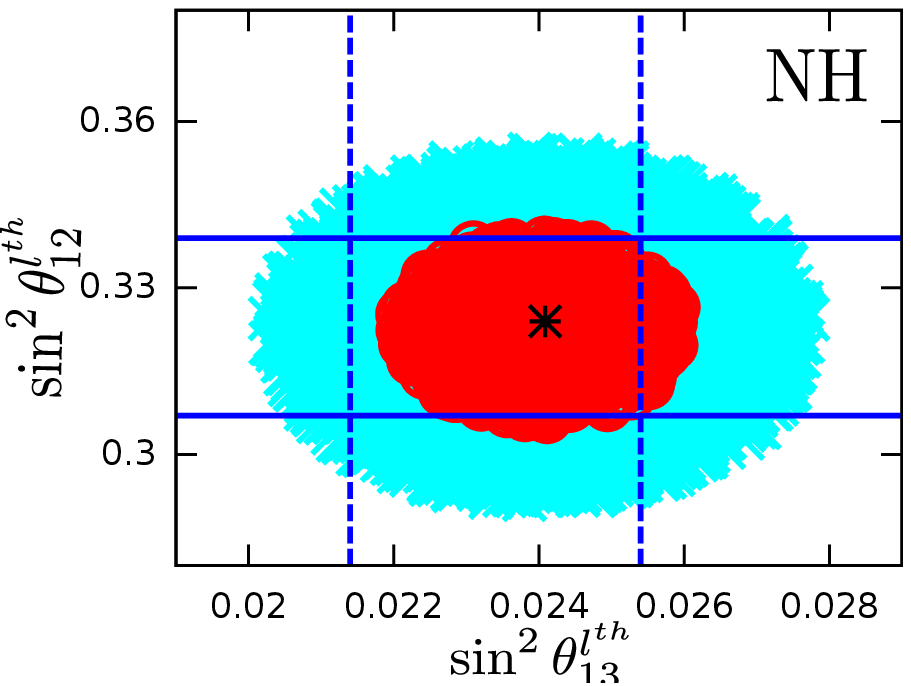}
  \includegraphics[scale=0.6]{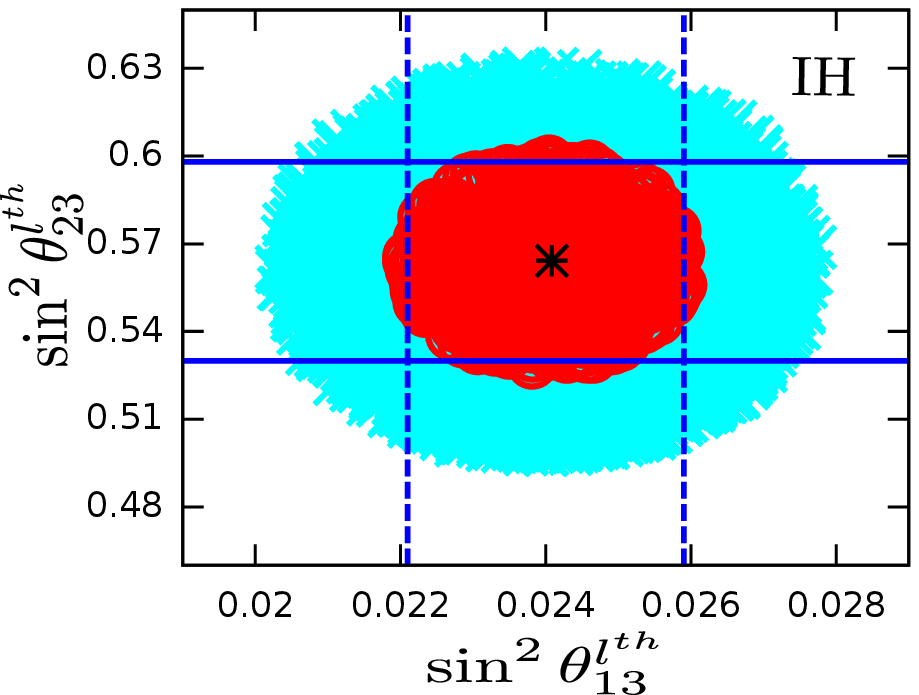}\hspace{6mm}\includegraphics[scale=0.6]{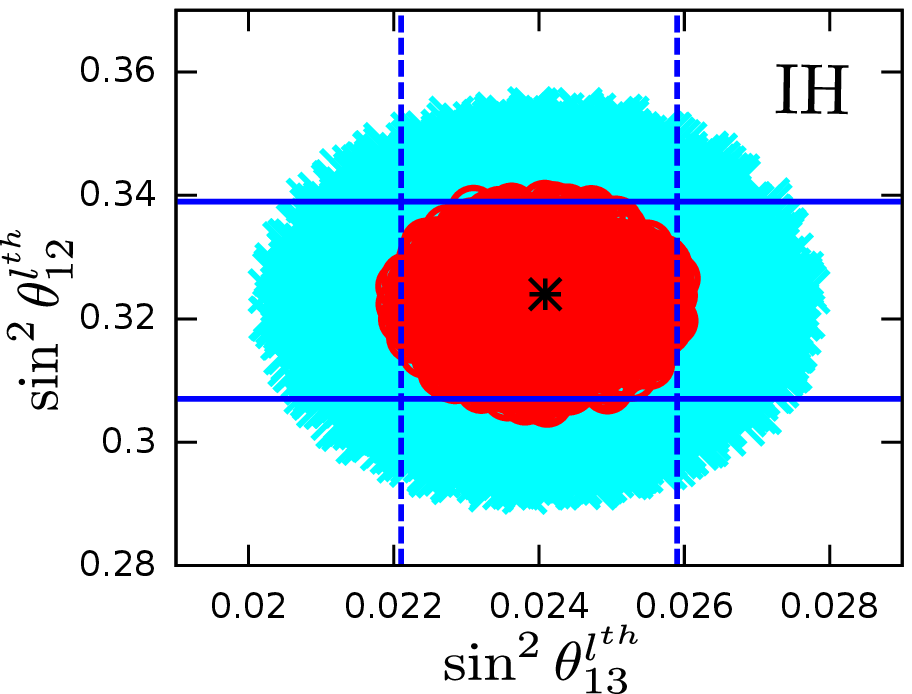}
  \caption{ Allowed regions for the lepton mixing angles considering a
    normal and inverted hierarchy (NH and IH) in the mass spectrum of
    neutrinos. The turquoise region is at $90\%$, the red region is at
    $69\%$. The blue solid lines delimit the experimental data of
    $\sin^{2} \theta_{12(23)}^{\ell}$ at $1\sigma$. The blue dashed
    lines delimit the experimental data of $\sin^{2}
    \theta_{13}^{\ell}$ at $1\sigma$~\cite{Forero:2014bxa}. }
  \label{eq29.4}
 \end{figure}

 Finally, as result of the $\chi^{2}$ analysis we obtained that for the best fit point 
 $\chi^{2}=0.020 [0.014]$ 
 for a normal [inverted] hierarchy. Also, the neutrino masses at $1~\sigma$ are:
{\scriptsize 
\begin{align}
  m_{\nu_{3}} = \left \{ \begin{array}{l} \vspace{2mm}
   \left( 5.35_{-1.73}^{+4.32} \right) \times 10^{-2}\textrm{eV}\\ 
   \left( 4.44_{-3.87}^{+4.21} \right) \times 10^{-2}\textrm{eV}
  \end{array}    \right. , \,
  m_{\nu_{2}} = \left \{ \begin{array}{l} \vspace{2mm}
   \left( 2.01_{-0.98}^{+6.42} \right) \times 10^{-2}\textrm{eV}\\ 
   \left( 6.71_{-1.81}^{+3.25} \right) \times 10^{-2}\textrm{eV}
  \end{array}    \right. , \,
  m_{\nu_{1}} = \left \{ \begin{array}{l} \vspace{2mm}
   \left( 1.08_{-1.30}^{+6.59} \right) \times 10^{-2}\textrm{eV}\\ 
   \left( 6.65_{-1.83}^{+3.27} \right) \times 10^{-2}\textrm{eV}
  \end{array}    \right. .
 \end{align}}
 The free parameters $\widetilde{\mu}_{0}$ and $\delta_{\nu}$ at $1~\sigma$ are:
 \begin{equation}
  \begin{array}{l}
    \widetilde{\mu}_{0} = 
    \left \{ \begin{array}{l} \vspace{2mm}
   0.22_{-0.20}^{+0.63} \\ 
   0.56_{-0.56}^{+0.29} 
  \end{array} \right.
  \quad \textrm{and} \quad 
  \delta_{\nu} = \left \{ \begin{array}{l} \vspace{2mm}
   0.75_{-0.15}^{+0.24} \\ 
   0.73_{-0.09}^{+0.25}
  \end{array} \right. .
  \end{array} 
 \end{equation}    
 We obtain the following numerical values for the leptonic mixing angles, at $1~\sigma$:
 \begin{equation}
  \begin{array}{l}
   \sin^{2} \theta_{12}^{\ell^{th}} = \left\{ \begin{array}{l} \vspace{2mm} 
    0.324_{-0.016}^{+0.015} \\ 
    0.325_{-0.018}^{+0.014}  
  \end{array} \right. ,
   \sin^{2} \theta_{23}^{\ell^{th}} = \left\{ \begin{array}{l} \vspace{2mm} 
     0.515_{-0.069}^{+0.078} \\ 
     0.562_{-0.048}^{+0.036} 
  \end{array} \right. ,
   \sin^{2} \theta_{13}^{\ell^{th}} = \left\{ \begin{array}{l} \vspace{2mm} 
    0.0232_{-0.0016}^{+0.0021} \\ 
    0.0241_{-0.0021}^{+0.0018}
  \end{array} \right. ,
 \end{array}
 \end{equation}  
 \begin{equation}
  \begin{array}{l}
    \theta_{12}^{\ell^{th}} = \left\{ \begin{array}{l} \vspace{2mm} 
     \left( 34.71_{-0.98}^{+0.91} \right)^{\circ} \\ 
     \left( 34.73_{-1.11}^{+0.89} \right)^{\circ} 
  \end{array} \right. ,
  \theta_{23}^{\ell^{th}} = \left\{ \begin{array}{l} \vspace{2mm} 
      \left( 45.83_{-3.98}^{+4.49} \right)^{\circ} \\ 
      \left( 48.57_{-2.76}^{+2.07} \right)^{\circ} 
  \end{array} \right. ,
  \theta_{13}^{\ell^{th}} = \left\{ \begin{array}{l} \vspace{2mm} 
       \left( 8.77_{-0.32}^{+0.40} \right)^{\circ} \\ 
       \left( 8.93_{-0.39}^{+0.33} \right)^{\circ} 
  \end{array} \right. ,
 \end{array}
 \end{equation}  
 which are in very good agreement with the last global fit reported in
 Ref~\cite{Forero:2014bxa}. The upper (lower) row corresponds to a normal
 (inverted) hierarchy. In the figure~\ref{eq29.4} we show the allowed
 regions, at $69\%$ and $95\%$ C.L., for the sine-squared of the lepton mixing
 angles considering a normal and inverted hierarchy.

   Before closing this section, a relevant comment is in
   order. As it is well known, the effective neutrino mass matrix,
   $\bf{M}_{\nu}$, is sensitive to the effect of the running of the
   mass matrix parameters from high to low energy  \cite{Babu:1993qv,
   Ray:2010rz}. Actually, small mixing angles could get a notable
   enhancement as was remarked in \cite{Babu:1993qv}.  Thus, in the
   scenario where the $\theta_{13}$ is tiny, it would be
   interesting to make a running of the mixing parameters to know how
   much the reactor angle changes, but we will leave this for a future
   work.
 
\section{Outlook and Remarks}

 We have studied a non-minimal supersymmetric $SU(5)$ model where the $Q_{6}$ 
 flavour symmetry plays an important role in accommodating the masses and 
 mixings for quarks an leptons. For the former sector, the CKM mixing matrix 
 has been obtained with great accuracy and it is consistent with the 
 experimental results. For leptons we considered two cases: $a)$~the first two 
 masses of RHN's are degenerate. $b)$~the RHN's masses are not degenerate. 
 
 On the one hand, for the case $a)$ the flavour 
 symmetry implies an appealing sum rule for the neutrino masses, which leads to an inverted 
 hierarchy and is crucial to determine the solar mixing angle.  As a main result in this case, 
 we have that the atmospheric $ \theta^{\ell^{th}}_{23} = \left( 46.18^{+0.66}_{-0.65} \right)^{\circ}$ 
 and solar 
 angle $\theta^{\ell^{th}}_{12} = \left( 36.62 \pm 4.06 \right)^{\circ}$ are in good agreement with 
 the experimental data, however, the reactor angle value, 
 $\theta^{\ell^{th}}_{13} = \left( 3.38^{+0.03}_{-0.02} \right)^{\circ}$, is more than $3 \sigma$ away 
 from the central value from the global fits. It is worth pointing out that 
 the model predicts a non-zero value for this angle, unlike the tri-bimaximal case,  and  
 that this value  constitutes the lower bound for $\theta_{13}$ in the more 
 general model, where the right-handed neutrinos in the $Q_6$ doublet are not 
 mass degenerate. 
 
On the other hand, for the more general case $b)$, where the RHN's masses are not degenerate, we obtained a value for the reactor mixing 
 angle in very good agreement with the last experimental data or global fits, as is the case in the 
 non-supersymmetric $S_3$ models~\cite{Canales:2012dr}. 
Namely, in this case it is possible that the neutrino mass spectrum 
  obeys a normal or an inverted hierarchy. Thus, we obtain that the leptonic mixing angles
  have the following theoretical values for a normal [inverted] hierarchy: 
  $\theta_{12}^{ \ell^{th} } = \left( 34.71_{-0.98}^{+0.91} \right)^{\circ}$ 
  $\left[ \left( 34.73_{-1.11}^{+0.89} \right)^{\circ} \right]$, 
  $\theta_{23}^{ \ell^{th} } = \left( 45.83_{-3.98}^{+4.49} \right)^{\circ}$ 
  $\left[ \left( 48.57_{-2.76}^{+2.07} \right)^{\circ} \right]$, and 
  $\theta_{13}^{ \ell^{th} } = \left( 8.77_{-0.32}^{+0.40} \right)^{\circ}$ 
  $\left[ \left( 8.93_{-0.39}^{+0.33} \right)^{\circ} \right] $.   

  Although in this preliminary analysis we have found the form of the
  mass and mixing matrices for quarks and leptons and we have shown
  that they lead to realistic values, we have left aside subtle issues
  as the full analysis of the scalar superpotential, 
    the details of the proton decay, the running of the mass
    parameters from high to low scale energy, and all the
  phenomenology that the model provides by itself. These topics will
  be taken into account in a complete study of the model, as we
  already commented. In general, the model seems to work out very well
  and it may be considered as a realistic one.
 
\section*{Acknowledgements}
We acknowledge useful discussions with P. Fileviez and
A. Mondrag\'on. This work was partially supported by the Mexican grants  PAPIIT 
IN113412 and IN111115, Conacyt  132059, the Spanish grants FPA2011-22975 and
Multidark CSD2009-00064 (MINECO), and PROMETEOII/2014/084 (Generalitat
Valenciana). JCGI thanks Red de Altas Energ\'{\i}as-CONACYT for the
financial support. FGC acknowledges the financial support from {\it
  CONACYT} and {\it PROMEP} under grants 208055 and 103.5/12/2548.

\appendix
\section{$Q_{6}$ Flavour symmetry}

 The $Q_{6}$ group has twelve elements which are contained in six conjugacy 
 classes, therefore, it contains six irreducible representations. We will use 
 the notation given in~\cite{Kajiyama:2007pr}, there are various notations and 
 extensive studies for this group, see for 
 example~\cite{Kubo:2012ty,Ishimori:2010au}. The $Q_{6}$ family symmetry has 
 ${\bf 2}$ two-dimensional irreducible representations denoted by 
 ${\bf 2}_{1}$ and ${\bf 2}_{2}$, $4$ one-dimensional ones which are denoted 
 by ${\bf 1}_{+,0}$, ${\bf 1}_{+,2}$, ${\bf 1}_{-,1}$ and ${\bf 1}_{-,3}$. As 
 it is well known, ${\bf 2}_{1}$ is a pseudo real and ${\bf 2}_{2}$ is a real 
 representation. In addition, for ${\bf 1}_{\pm,n}$ we have that $n=0,1,2,3$ 
 is the factor $\exp\left(in\pi/2\right)$ that appears in the matrix given by 
 ${\bf B}$. The $\pm$ stands for the change of sign under the transformation 
 given by the ${\bf A}$ matrix. So that the first two one-dimensional 
 representations are real and the two latter ones are complex conjugate to 
 each other.
 \begin{equation}
  Q_{6} = \lbrace 1, {\bf A}, {\bf A}^{2}, {\bf A}^{3}, {\bf A}^{4}, 
   {\bf A}^{5}, {\bf B}, {\bf A}{\bf B}, {\bf A}^{2}{\bf B}, {\bf A}^{3} 
   {\bf B}, {\bf A}^{4} {\bf B}, {\bf A}^{5} {\bf B} \rbrace,
 \end{equation}
 where the ${\bf A}$ and ${\bf B}$ are two-dimensional matrices whose explicit 
 forms are given by
 \begin{equation}
  {\bf A} = 
  \begin{pmatrix} 
   \cos\left(\pi/3 \right) & \sin\left(\pi/3 \right) \\ 
   -\sin\left(\pi/3 \right) & \cos\left(\pi/3 \right)
  \end{pmatrix}
  \quad \textrm{and} \quad
  {\bf B} = 
  \begin{pmatrix}
   i & 0 \\ 
   0 & -i
  \end{pmatrix} .
 \end{equation}
 Let us write the multiplication rules among the six irreducible 
 representations which will be useful to build a phenomenological model: 
 {\scriptsize
 \begin{align}\label{q6rul}
  {\bf 1}_{+,2}\otimes{\bf 1}_{+,2} & =  {\bf 1}_{+,0},\quad {\bf 1}_{-,3} \otimes {\bf 1}_{-,3} 
  = {\bf 1}_{+,2},\quad {\bf 1}_{-,1} \otimes {\bf 1}_{-,1} = {\bf 1}_{+,2},\quad {\bf 1}_{-,1} 
  \otimes {\bf 1}_{-,3} = {\bf 1}_{+,0},\nn\\
  {\bf 1}_{+,2} \otimes {\bf 1}_{-,1} &=  {\bf 1}_{-,3},\quad {\bf 1}_{+,2} \otimes {\bf 1}_{-,3} 
  = {\bf 1}_{-,1},\quad {\bf 2}_{1} \otimes  {\bf 1}_{+,2} = {\bf 2}_{1},\quad {\bf 2}_{1} \otimes 
  {\bf 1}_{-,3} = {\bf 2}_{2},\nn\\
  {\bf 2}_{1} \otimes {\bf 1}_{-,1} &=  {\bf 2}_{2},\quad {\bf 2}_{2} \otimes {\bf 1}_{+,2} = {\bf 2}_{2},\quad
   {\bf 2}_{2} \otimes {\bf 1}_{-,3} = {\bf 2}_{1},\quad {\bf 2}_{2} \otimes {\bf 1}_{-,1} = {\bf 2}_{1}; \nn\\
\overbrace{\begin{pmatrix}
   x_{1} \\ 
   x_{2}
  \end{pmatrix}}^{{\bf 2}_{1}}
   \, \otimes \, 
 \overbrace{\begin{pmatrix}
   y_{1} \\ 
   y_{2}
  \end{pmatrix}}^{{\bf 2}_{1}} & =  
 \overbrace{\left(x_{1}y_{2}-x_{2}y_{1}\right)}^{{\bf 1}_{+,0}} \quad +\quad \overbrace{\left(x_{1}y_{1}+x_{2}y_{2}\right)}^{{\bf 1}_{+,2}} \quad + \quad
  \overbrace{\begin{pmatrix}
   -x_{1}y_{2}-x_{2}y_{1} \\ 
   x_{1}y_{1}-x_{2}y_{2}
  \end{pmatrix}}^{{\bf 2}_{2}} \nn\\
  \overbrace{\begin{pmatrix}
   a_{1} \\ 
   a_{2}
  \end{pmatrix}}^{{\bf 2}_{2}} \otimes 
 \overbrace{\begin{pmatrix}
   b_{1} \\ 
   b_{2}
  \end{pmatrix}}^{{\bf 2}_{2}}&=\overbrace{\left(a_{1}b_{1}+a_{2}b_{2}\right)}^{{\bf 1}_{+,0}}\quad + \quad
 \overbrace{\left(a_{1}b_{2}-  a_{2}b_{1}\right)}^{{\bf 1}_{+,2}} \quad + \quad
 \overbrace{\begin{pmatrix}
   -a_{1}b_{1}+a_{2}b_{2} \\ 
   a_{1}b_{2}+a_{2}b_{1}
  \end{pmatrix}}^{{\bf 2}_{2}} \nn\\
  \overbrace{\begin{pmatrix}
   x_{1} \\ 
   x_{2}
  \end{pmatrix}}^{{\bf 2}_{1}} \otimes 
  \overbrace{\begin{pmatrix}
   a_{1} \\ 
   a_{2}
  \end{pmatrix}}^{{\bf 2}_{2}} &=\overbrace{\left(x_{1}a_{2}+x_{2}a_{1}\right)}^{{\bf 1}_{-,3}}\quad + \quad  
  \overbrace{\left(x_{1}a_{1}-x_{2}a_{2}\right)}^{{\bf 1}_{-,1}} \quad + \quad
 \overbrace{\begin{pmatrix}
   x_{1}a_{1}+x_{2}a_{2} \\ 
   x_{1}a_{2}-x_{2}a_{1}
  \end{pmatrix}}^{{\bf 2}_{1}},
 \end{align}}

\section{Neutrino Mass Matrix}\label{ApenC}
 The degeneracy in the vacuum expectation values for two scalar fields, does not 
 modify the functional structure of the effective neutrino mass matrix 
 ${\bf m}_{\nu}$, eq.~(\ref{ML:ec:13}), because it still is a matrix with one 
 texture zero. But the parameters $a_{\nu}$, $b_{\nu}$, $c_{\nu}$, $d_{\nu}$
 and $\mu_{0}$ given in eq.~(\ref{ML:ec:14}) are simplified a little. 
 Therefore, the expressions in eqs~(\ref{ML:ec:13}) and~(\ref{ML:ec:14}) take 
 the following form 
 \begin{equation}\label{Ap:ec:13}
  {\bf m}_{\nu} = {\bf \textit{u} }_{\theta}^{T} {\bf M}_{\nu} 
  {\bf \textit{u} }_{\theta} =
  \left( \begin{array}{ccc}
   b_{\nu} & a_{\nu} & c_{\nu} \\
   a_{\nu} & \mu_{0} & 0       \\
   c_{\nu} & 0       & d_{\nu}
  \end{array}   \right) 
  \quad \textrm{and} \quad 
  {\bf \textit{u} }_{\theta} =
  \left( \begin{array}{ccc}
   \cos \theta & 0 & -\sin \theta \\
   \sin \theta & 0 &  \cos \theta \\
   0           & 1 & 0
  \end{array}   \right),
 \end{equation}  
 with  
 \begin{equation}\label{Ap:ec:14}
  \begin{array}{l} \vspace{2mm}
  \tan \theta = \frac{ M_{R_{2}} }{ M_{R_{1}} }, \quad
  \mu_{0} = \frac{ \left( y_{2}^{n} h^{0 u} \right)^{2} }{ 
   M_{R_{1}} M_{R_{2}} } 
  \left[ M_{R_{2}} + M_{R_{1}} \right] , \quad
  a_{\nu} = \frac{ \cos \theta y_{1}^{n} h^{0 u}_{3} y_{2}^{n} h^{0 u} }{ 
  M_{R_{2}} M_{R_{1}}^{2}  } 
  \left( M_{R_{1}}^{2}  + M_{R_{2}}^{2} \right), \\ \vspace{2mm}
  b_{\nu} = \frac{\cos^{2} \theta}{ M_{R_{1}}^{2} } 
   \left[ \frac{ \left( y_{1}^{n} h^{0 u}_{3} \right)^{2} }{ M_{R_{1}} 
   M_{R_{2} }  } 
  \left( M_{R_{1}}^{3} + M_{R_{2}}^{3}  \right) 
   + \frac{ \left( y_{3}^{n} h^{0 u} \right)^{2} }{ M_{R_{3}} } \left( 
   M_{R_{1}} + M_{R_{2}} 
   \right)^{2} \right], \\ \vspace{2mm}
  c_{\nu} =  \frac{  \cos^{2} \theta }{ M_{R_{1}}^{2} } 
   \left[ \left( y_{1}^{n} h^{0 u}_{3} \right)^{2} \left( M_{R_{2}} - 
   M_{R_{1}} \right) 
   + \frac{ \left( y_{3}^{n}  h^{0 u} \right)^{2} }{ M_{R_{3}} } 
    \left( M_{R_{1}}^{2} -  M_{R_{2}}^{2} \right)  \right],
   \; \textrm{and} \;
  \\ \vspace{2mm} 
  d_{\nu} = \frac{ \cos^{2} \theta }{ \left( M_{R_{1}} \right)^{2} } 
   \left[ \left( y_{1}^{n} h^{0 u}_{3} \right)^{2} \left( M_{R_{2}}  +  
   M_{R_{1}} \right) 
   + \frac{ \left( y_{3}^{n} h^{0 u} \right)^{2} }{ M_{R_{3}} } 
   \left( M_{R_{1}} - M_{R_{2}} \right)^{2} \right].
  \end{array}  
 \end{equation} 
 From the previous expressions is easy to notice that when the masses of the first 
 two right-handed neutrinos are degenerate, $M_{R_{1}} = M_{R_{2}}$, the 
 parameter $c_{\nu}$ becomes zero and the rotation angle takes the value 
 $\theta = \pi/4$. Hence, the effective neutrino mass matrix ${\bf m}_{\nu}$ 
 is reduced to a block matrix. 

\bibliographystyle{unsrt}
\bibliography{references.bib}

\begin{thebibliography}{10}

\bibitem{Masiero:2005ua}
A.~Masiero, S.~K. Vempati, and O.~Vives.
\newblock {Flavour physics and grand unification}.
\newblock {\em 0711.2903}, 2005.

\bibitem{Fritzsch:1977za}
H.~Fritzsch.
\newblock {Calculating the Cabibbo Angle}.
\newblock {\em Phys.Lett.}, B70:436, 1977.

\bibitem{Fritzsch:1977vd}
H.~Fritzsch.
\newblock {Weak Interaction Mixing in the Six - Quark Theory}.
\newblock {\em Phys.Lett.}, B73:317--322, 1978.

\bibitem{Kobayashi:1973fv}
Makoto Kobayashi and Toshihide Maskawa.
\newblock {CP Violation in the Renormalizable Theory of Weak Interaction}.
\newblock {\em Prog. Theor. Phys.}, 49:652--657, 1973.

\bibitem{Cabibbo:1963yz}
Nicola Cabibbo.
\newblock {Unitary Symmetry and Leptonic Decays}.
\newblock {\em Phys.Rev.Lett.}, 10:531--533, 1963.

\bibitem{Branco:2010tx}
G.~C. Branco, D.~Emmanuel-Costa, and C.~Simoes.
\newblock {Nearest-Neighbour Interaction from an Abelian Symmetry and
  Deviations from Hermiticity}.
\newblock {\em Phys. Lett.}, B690:62--67, 2010.

\bibitem{Fritzsch:2011cu}
Harald Fritzsch, Zhi-zhong Xing, and Ye-Ling Zhou.
\newblock {Non-Hermitian Perturbations to the Fritzsch Textures of Lepton and
  Quark Mass Matrices}.
\newblock {\em Phys. Lett.}, B697:357--363, 2011.

\bibitem{Maki:1962mu}
Ziro Maki, Masami Nakagawa, and Shoichi Sakata.
\newblock {Remarks on the unified model of elementary particles}.
\newblock {\em Prog.Theor.Phys.}, 28:870--880, 1962.

\bibitem{Pontecorvo:1967fh}
B.~Pontecorvo.
\newblock {Neutrino experiments and the question of leptonic-charge
  conservation}.
\newblock {\em Sov. Phys. JETP}, 26:984--988, 1968.

\bibitem{Harayama:1996am}
K.~Harayama and N.~Okamura.
\newblock {Exact parametrization of the mass matrices and the KM matrix}.
\newblock {\em Phys.Lett.}, B387:614--622, 1996.

\bibitem{Harayama:1996jr}
K.~Harayama, N.~Okamura, A.I. Sanda, and Zhi-Zhong Xing.
\newblock {Getting at the quark mass matrices}.
\newblock {\em Prog.Theor.Phys.}, 97:781--790, 1997.

\bibitem{Ishimori:2010au}
Hajime Ishimori, Tatsuo Kobayashi, Hiroshi Ohki, Yusuke Shimizu, Hiroshi Okada,
  and Morimitsu Tanimoto.
\newblock Non-abelian discrete symmetries in particle physics.
\newblock {\em Progress of Theoretical Physics Supplement}, 183:1--163, 2010.

\bibitem{Babu:2004tn}
Kaladi~S. Babu and Jisuke Kubo.
\newblock {Dihedral families of quarks, leptons and Higgses}.
\newblock {\em Phys.Rev.}, D71:056006, 2005.

\bibitem{Kajiyama:2005rk}
Yuji Kajiyama, Etsuko Itou, and Jisuke Kubo.
\newblock {Nonabelian discrete family symmetry to soften the SUSY flavor
  problem and to suppress proton decay}.
\newblock {\em Nucl. Phys.}, B743:74--103, 2006.

\bibitem{Kajiyama:2007pr}
Yuji Kajiyama.
\newblock {R-parity violation and non-Abelian discrete family symmetry}.
\newblock {\em JHEP}, 04:007, 2007.

\bibitem{Babu:2009nn}
K.S. Babu and Yanzhi Meng.
\newblock {Flavor Violation in Supersymmetric Q(6) Model}.
\newblock {\em Phys.Rev.}, D80:075003, 2009.

\bibitem{Babu:2011mv}
K.S. Babu, Kenji Kawashima, and Jisuke Kubo.
\newblock {Variations on the Supersymmetric $Q_6$ Model of Flavor}.
\newblock {\em Phys.Rev.}, D83:095008, 2011.

\bibitem{Georgi:1974yf}
H.~Georgi, Helen~R. Quinn, and Steven Weinberg.
\newblock {Hierarchy of Interactions in Unified Gauge Theories}.
\newblock {\em Phys.Rev.Lett.}, 33:451--454, 1974.

\bibitem{Pati:1974yy}
Jogesh~C. Pati and Abdus Salam.
\newblock {Lepton Number as the Fourth Color}.
\newblock {\em Phys.Rev.}, D10:275--289, 1974.

\bibitem{Mohapatra:1974gc}
R.N. Mohapatra and Jogesh~C. Pati.
\newblock {A Natural Left-Right Symmetry}.
\newblock {\em Phys.Rev.}, D11:2558, 1975.

\bibitem{Georgi:1974sy}
H.~Georgi and S.L. Glashow.
\newblock {Unity of All Elementary Particle Forces}.
\newblock {\em Phys.Rev.Lett.}, 32:438--441, 1974.

\bibitem{Langacker:1980js}
Paul Langacker.
\newblock {Grand Unified Theories and Proton Decay}.
\newblock {\em Phys. Rept.}, 72:185, 1981.

\bibitem{Buras:1977yy}
A.~J. Buras, John~R. Ellis, M.~K. Gaillard, and Dimitri~V. Nanopoulos.
\newblock {Aspects of the Grand Unification of Strong, Weak and Electromagnetic
  Interactions}.
\newblock {\em Nucl. Phys.}, B135:66--92, 1978.

\bibitem{Dimopoulos:1981zb}
Savas Dimopoulos and Howard Georgi.
\newblock {Softly Broken Supersymmetry and SU(5)}.
\newblock {\em Nucl.Phys.}, B193:150, 1981.

\bibitem{Dimopoulos:1981yj}
S.~Dimopoulos, S.~Raby, and Frank Wilczek.
\newblock {Supersymmetry and the Scale of Unification}.
\newblock {\em Phys.Rev.}, D24:1681--1683, 1981.

\bibitem{Sakai:1981gr}
N.~Sakai.
\newblock {Naturalness in Supersymmetric Guts}.
\newblock {\em Z.Phys.}, C11:153, 1981.

\bibitem{Sakai:1981pk}
N.~Sakai and Tsutomu Yanagida.
\newblock {Proton Decay in a Class of Supersymmetric Grand Unified Models}.
\newblock {\em Nucl.Phys.}, B197:533, 1982.

\bibitem{Weinberg:1981wj}
Steven Weinberg.
\newblock {Supersymmetry at Ordinary Energies. 1. Masses and Conservation
  Laws}.
\newblock {\em Phys.Rev.}, D26:287, 1982.

\bibitem{Ellis:1981tv}
John~R. Ellis, Dimitri~V. Nanopoulos, and Serge Rudaz.
\newblock {GUTs 3: SUSY GUTs 2}.
\newblock {\em Nucl. Phys.}, B202:43, 1982.

\bibitem{Dimopoulos:1981dw}
Savas Dimopoulos, Stuart Raby, and Frank Wilczek.
\newblock {Proton Decay in Supersymmetric Models}.
\newblock {\em Phys.Lett.}, B112:133, 1982.

\bibitem{Nath:1985ub}
Pran Nath, Ali~H. Chamseddine, and Richard~L. Arnowitt.
\newblock {Nucleon Decay in Supergravity Unified Theories}.
\newblock {\em Phys.Rev.}, D32:2348--2358, 1985.

\bibitem{Hisano:1992jj}
J.~Hisano, H.~Murayama, and T.~Yanagida.
\newblock {Nucleon decay in the minimal supersymmetric SU(5) grand
  unification}.
\newblock {\em Nucl.Phys.}, B402:46--84, 1993.

\bibitem{Goto:1998qg}
Toru Goto and Takeshi Nihei.
\newblock {Effect of RRRR dimension five operator on the proton decay in the
  minimal SU(5) SUGRA GUT model}.
\newblock {\em Phys.Rev.}, D59:115009, 1999.

\bibitem{Murayama:2001ur}
Hitoshi Murayama and Aaron Pierce.
\newblock {Not even decoupling can save minimal supersymmetric SU(5)}.
\newblock {\em Phys.Rev.}, D65:055009, 2002.

\bibitem{Bajc:2002bv}
Borut Bajc, Pavel Fileviez~Perez, and Goran Senjanovic.
\newblock {Proton decay in minimal supersymmetric SU(5)}.
\newblock {\em Phys.Rev.}, D66:075005, 2002.

\bibitem{Martens:2010nm}
W.~Martens, L.~Mihaila, J.~Salomon, and M.~Steinhauser.
\newblock {Minimal Supersymmetric SU(5) and Gauge Coupling Unification at Three
  Loops}.
\newblock {\em Phys.Rev.}, D82:095013, 2010.

\bibitem{Hisano:2013exa}
Junji Hisano, Daiki Kobayashi, Takumi Kuwahara, and Natsumi Nagata.
\newblock {Decoupling Can Revive Minimal Supersymmetric SU(5)}.
\newblock {\em JHEP}, 1307:038, 2013.

\bibitem{Romao:1991ex}
J.~C. Romao and J.~W.~F. Valle.
\newblock {Neutrino masses in supersymmetry with spontaneously broken R
  parity}.
\newblock {\em Nucl. Phys.}, B381:87--108, 1992.

\bibitem{Hirsch:2000ef}
M.~Hirsch, M.A. Diaz, W.~Porod, J.C. Romao, and J.W.F. Valle.
\newblock {Neutrino masses and mixings from supersymmetry with bilinear R
  parity violation: A Theory for solar and atmospheric neutrino oscillations}.
\newblock {\em Phys.Rev.}, D62:113008, 2000.

\bibitem{GellMann:1980vs}
Murray Gell-Mann, Pierre Ramond, and Richard Slansky.
\newblock {Complex Spinors and Unified Theories}.
\newblock {\em Conf.Proc.}, C790927:315--321, 1979.

\bibitem{fukugita2003physics}
M.~Fukugita and T.~Yanagida.
\newblock {\em Physics of Neutrinos: And Applications to Astrophysics}.
\newblock Physics and astronomy online library. Springer, 2003.

\bibitem{Mohapatra:1980yp}
Rabindra~N. Mohapatra and Goran Senjanovic.
\newblock {Neutrino Masses and Mixings in Gauge Models with Spontaneous Parity
  Violation}.
\newblock {\em Phys. Rev.}, D23:165, 1981.

\bibitem{Mohapatra:1979ia}
Rabindra~N. Mohapatra and Goran Senjanovic.
\newblock {Neutrino Mass and Spontaneous Parity Violation}.
\newblock {\em Phys.Rev.Lett.}, 44:912, 1980.

\bibitem{Minkowski:1977sc}
Peter Minkowski.
\newblock {mu $\to$ e gamma at a Rate of One Out of 1-Billion Muon Decays?}
\newblock {\em Phys. Lett.}, B67:421, 1977.

\bibitem{Ishimori:2008fi}
Hajime Ishimori, Yusuke Shimizu, and Morimitsu Tanimoto.
\newblock {S4 Flavor Symmetry of Quarks and Leptons in SU(5) GUT}.
\newblock {\em Prog. Theor. Phys.}, 121:769--787, 2009.

\bibitem{Hagedorn:2010th}
Claudia Hagedorn, Stephen~F. King, and Christoph Luhn.
\newblock {A SUSY GUT of Flavour with S4 x SU(5) to NLO}.
\newblock {\em JHEP}, 06:048, 2010.

\bibitem{Chen:2013wba}
Mu-Chun Chen, Jinrui Huang, K.T. Mahanthappa, and Alexander~M. Wijangco.
\newblock {Large $\theta_{13}$ in a SUSY SU(5) x T' Model}.
\newblock {\em JHEP}, 1310:112, 2013.

\bibitem{Felix:2006pn}
O.~Felix, A.~Mondragon, M.~Mondragon, and E.~Peinado.
\newblock {Neutrino masses and mixings in a minimal S(3)-invariant extension of
  the standard model}.
\newblock {\em AIP Conf. Proc.}, 917:383--389, 2007.

\bibitem{Mondragon:2007af}
A.~Mondragon, M.~Mondragon, and E.~Peinado.
\newblock {Lepton masses, mixings and FCNC in a minimal $S_3$-invariant
  extension of the Standard Model}.
\newblock {\em Phys. Rev.}, D76:076003, 2007.

\bibitem{Canales:2012dr}
F.~Gonzalez~Canales, A.~Mondragon, and M.~Mondragon.
\newblock {The $S_3$ Flavour Symmetry: Neutrino Masses and Mixings}.
\newblock {\em Fortsch.Phys.}, 61:546--570, 2013.

\bibitem{Einhorn:1981sx}
M.B. Einhorn and D.R.T. Jones.
\newblock {The Weak Mixing Angle and Unification Mass in Supersymmetric SU(5)}.
\newblock {\em Nucl.Phys.}, B196:475, 1982.

\bibitem{Astorga:1994gh}
F.~Astorga.
\newblock {Constraints from unification in SU(5) and SUSY SU(5)}.
\newblock {\em J.Phys.}, G20:241--260, 1994.

\bibitem{Berezhiani:1998hg}
Z.~Berezhiani, Z.~Tavartkiladze, and M.~Vysotsky.
\newblock {d = 5 operators in SUSY GUT: Fermion masses versus proton decay}.
\newblock {\em arXiv:hep-ph/9809301}, 1998.

\bibitem{Bajc:2002pg}
Borut Bajc, Pavel Fileviez~Perez, and Goran Senjanovic.
\newblock Minimal supersymmetric su(5) theory and proton decay: Where do we
  stand?
\newblock pages 131--139, 2002.

\bibitem{Dorsner:2007fy}
Ilja Dorsner and Irina Mocioiu.
\newblock {Predictions from type II see-saw mechanism in SU(5)}.
\newblock {\em Nucl.Phys.}, B796:123--136, 2008.

\bibitem{Fritzsch:1999ee}
Harald Fritzsch and Zhi-zhong Xing.
\newblock {Mass and flavor mixing schemes of quarks and leptons}.
\newblock {\em Prog. Part. Nucl. Phys.}, 45:1--81, 2000.

\bibitem{Canales:2013ura}
F.~González Canales, A.~Mondragón, M.~Mondragón, U.J. Saldaña~Salazar, and
  L.~Velasco-Sevilla.
\newblock {Fermion mixing in an $S_{3}$ model with three Higgs doublets}.
\newblock {\em J.Phys.Conf.Ser.}, 447:012053, 2013.

\bibitem{GonzalezCanales:2012kj}
F.~Gonzalez~Canales and A.~Mondragon.
\newblock {The neutrino mixing angle theta(13) in an S(3) flavour symmetric
  model}.
\newblock {\em J.Phys.Conf.Ser.}, 387:012008, 2012.

\bibitem{GonzalezCanales:2012za}
F.~Gonzalez~Canales and A.~Mondragon.
\newblock {The flavour symmetry S(3) and the neutrino mass matrix with two
  texture zeroes}.
\newblock {\em J.Phys.Conf.Ser.}, 378:012014, 2012.

\bibitem{Canales:2013cga}
F.~González~Canales, A.~Mondragón, M.~Mondragón, U.~J. Saldaña~Salazar, and
  L.~Velasco-Sevilla.
\newblock {Quark sector of S3 models: classification and comparison with
  experimental data}.
\newblock {\em Phys.Rev.}, D88:096004, 2013.

\bibitem{Canales:2011ug}
F.~Gonzalez~Canales and A.~Mondragon.
\newblock {The $S_{3}$ symmetry: Flavour and texture zeroes}.
\newblock {\em J. Phys. Conf. Ser.}, 287:012015, 2011.

\bibitem{Nakamura:2010zzi}
K.~Nakamura et~al.
\newblock {Review of particle physics}.
\newblock {\em J. Phys.}, G37:075021, 2010.

\bibitem{Schwetz:2011qt}
Thomas Schwetz, Mariam Tortola, and J.W.F. Valle.
\newblock {Global neutrino data and recent reactor fluxes: status of
  three-flavour oscillation parameters}.
\newblock {\em New J.Phys.}, 13:063004, 2011.
\newblock * Temporary entry *.

\bibitem{Adamson:2011qu}
P.~Adamson et~al.
\newblock {Improved search for muon-neutrino to electron-neutrino oscillations
  in MINOS}.
\newblock {\em Phys.Rev.Lett.}, 107:181802, 2011.

\bibitem{Beringer:1900zz}
J.~Beringer et~al.
\newblock {Review of Particle Physics (RPP)}.
\newblock {\em Phys.Rev.}, D86:010001, 2012.

\bibitem{Xing:2007fb}
Zhi-zhong Xing, He~Zhang, and Shun Zhou.
\newblock {Updated Values of Running Quark and Lepton Masses}.
\newblock {\em Phys.Rev.}, D77:113016, 2008.

\bibitem{Canales:2012ix}
F.~Gonzalez Canales, A.~Mondragon, U.J.~Saldana Salazar, and
  L.~Velasco-Sevilla.
\newblock {$S_3$ as a unified family theory for quarks and leptons}.
\newblock {\em arXiv:1210.0288}, 2012.

\bibitem{Tortola:2012te}
D.V. Forero, M.~Tortola, and J.W.F. Valle.
\newblock {Global status of neutrino oscillation parameters after
  Neutrino-2012}.
\newblock {\em Phys.Rev.}, D86:073012, 2012.

\bibitem{GonzalezGarcia:2012sz}
M.C. Gonzalez-Garcia, Michele Maltoni, Jordi Salvado, and Thomas Schwetz.
\newblock {Global fit to three neutrino mixing: critical look at present
  precision}.
\newblock {\em JHEP}, 1212:123, 2012.

\bibitem{Forero:2014bxa}
D.V. Forero, M.~Tortola, and J.W.F. Valle.
\newblock {Neutrino oscillations refitted}.
\newblock {\em arXiv:1405.7540}, 2014.

\bibitem{Ade:2013zuv}
P.A.R. Ade et~al.
\newblock {Planck 2013 results. XVI. Cosmological parameters}.
\newblock {\em Astron.Astrophys.}, 2014.

\bibitem{Babu:1993qv}
K.S. Babu, Chung~Ngoc Leung, and James~T. Pantaleone.
\newblock {Renormalization of the neutrino mass operator}.
\newblock {\em Phys.Lett.}, B319:191--198, 1993.

\bibitem{Ray:2010rz}
Shamayita Ray.
\newblock {Renormalization group evolution of neutrino masses and mixing in
  seesaw models: A Review}.
\newblock {\em Int.J.Mod.Phys.}, A25:4339--4384, 2010.

\bibitem{Kubo:2012ty}
Jisuke Kubo.
\newblock {Super Flavorsymmetry with Multiple Higgs Doublets}.
\newblock {\em Fortsch.Phys.}, 61:597--621, 2013.

\end{thebibliography}
\end{document}